%% file: cas-dc-sample.tex
\documentclass[a4paper,fleqn]{cas-dc}

\usepackage[numbers]{natbib}
\usepackage{amsmath}
\usepackage{hyperref}
\def\tsc#1{\csdef{#1}{\textsc{\lowercase{#1}}\xspace}}
\tsc{WGM}
\tsc{QE}
\tsc{EP}
\tsc{PMS}
\tsc{BEC}
\tsc{DE}

\begin{document}
\let\WriteBookmarks\relax
\def\floatpagepagefraction{1}
\def\textpagefraction{.001}
\shorttitle{}
\shortauthors{M. Tahir et~al.}

\title [mode = title]{Artificial Intelligence and Deep Learning Algorithms for Epigenetic Sequence Analysis: A Review for Epigeneticists and AI Experts}                      



\author[1]{Muhammad Tahir}[
                        ]

\credit{Conceived the idea, analyzed the studies and wrote this manuscript}

\affiliation[1]{organization={Department of Electrical and Computer Engineering, University of Manitoba},
                city={Winnipeg},
                postcode={R3T 5V6}, 
                state={MB},
                country={Canada}}

\author[1]{Mahboobeh Norouzi}[
                        ]

\credit{Analyzed the studies and wrote this manuscript}


\author[2]{Shehroz S. Khan}[
                        ]

\credit{Conceived the idea, analyzed the studies and wrote this manuscript}

\affiliation[2]{organization={KITE, University Health Network},
                city={Toronto},
                country={Canada}}

\author[3]{James R. Davie}[
                        ]

\credit{Conceived the idea, analyzed the studies and wrote this manuscript}

\affiliation[3]{organization={Department of Biochemistry and Medical Genetics, Max Rady College of Medicine, Rady Faculty of Health Sciences, University of Manitoba},
                city={Winnipeg},
                state={MB},
                country={Canada}}

\author[4]{Soichiro Yamanaka}[
                        ]

\credit{Conceived the idea, analyzed the studies and wrote this manuscript}

\affiliation[4]{organization={Graduate School of Science, Department of Biophysics and Biochemistry, University of Tokyo},
                country={Japan}}

\author[1]{Ahmed Ashraf}
\cormark[1]
\ead{ahmed.ashraf@umanitoba.ca}

\credit{Conceived the idea, analyzed the studies and wrote this manuscript}

\cortext[cor1]{Corresponding author}


\begin{abstract}
Epigenetics encompasses mechanisms that can alter the expression of genes without changing the underlying genetic sequence. The epigenetic regulation of gene expression is initiated and sustained by several mechanisms such as DNA methylation, histone modifications, chromatin conformation, and non-coding RNA. The changes in gene regulation and expression can manifest in the form of various diseases and disorders such as cancer and congenital deformities. Over the last few decades, high-throughput experimental approaches have been used to identify and understand epigenetic changes, but these laboratory experimental approaches and biochemical processes are time-consuming and expensive. To overcome these challenges, machine learning and artificial intelligence (AI) approaches have been extensively used for mapping epigenetic modifications to their phenotypic manifestations. In this paper we provide a narrative review of published research on AI models trained on epigenomic data to address a variety of problems such as prediction of disease markers, gene expression, enhancer-promoter interaction, and chromatin states. The purpose of this review is twofold as it is addressed to both AI experts and epigeneticists. For AI researchers, we provided a taxonomy of epigenetics research problems that can benefit from an AI-based approach. For epigeneticists, given each of the above problems we provide a list of candidate AI solutions in the literature. We have also identified several gaps in the literature, research challenges, and recommendations to address these challenges.
\end{abstract}

\begin{graphicalabstract}
    \centering
    \includegraphics[width=0.99\textwidth]{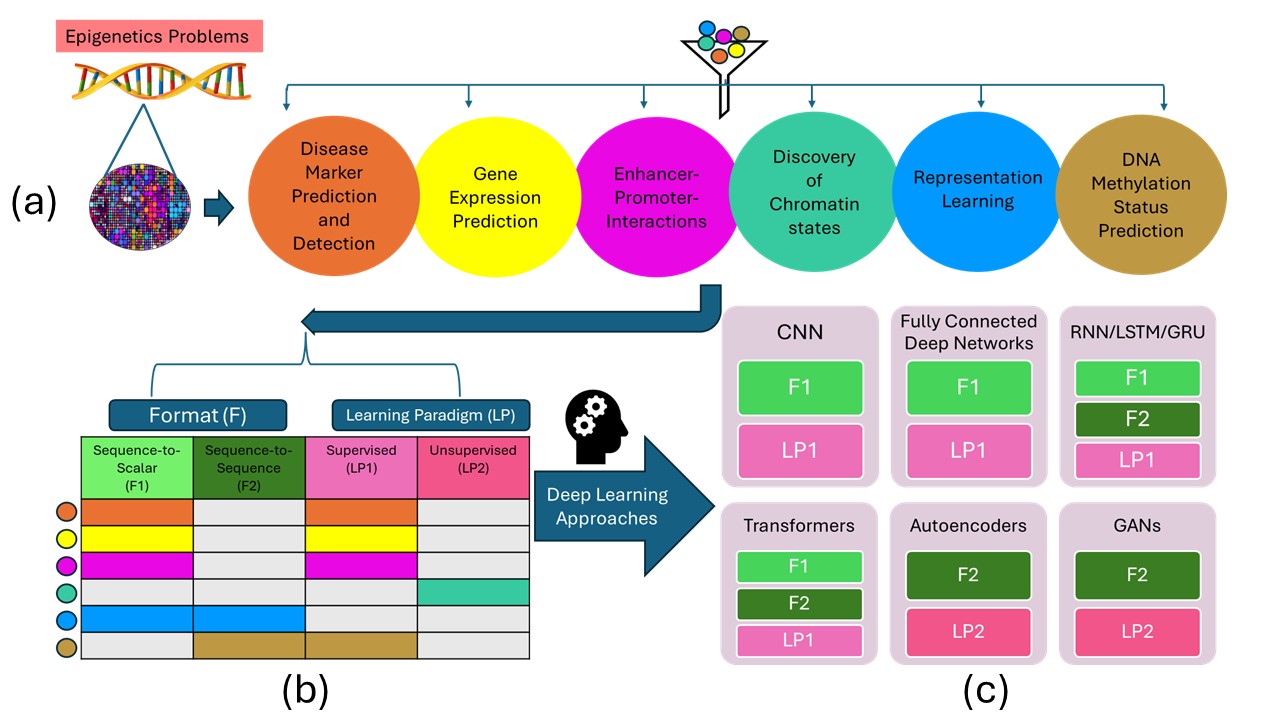}

    \vspace{0.5cm}  

    \textbf{Caption:} Graphical overview of the taxonomy of various problems in epigenetic sequence analysis as mapped to AI-based solutions available in the literature. This review article is intended toward Epigeneticists and AI experts. (a) Different types of epigenetic problems. (b) A tabular illustration of which problem corresponds to which format (F) and learning paradigm (LP), e.g., first row shows that ‘Disease marker prediction and detection’ (orange) corresponds to Format 1 (F1: Sequence-to-calar) and Learning Paradigm 1 (LP1: Supervised). (c) Illustration of which neural network architecture corresponds to which F and LP, e.g., Autoencoders in the context of sequential inputs would correspond to F2: Sequence-to-sequence and LP2: Unsupervised.
\end{graphicalabstract}

\begin{highlights}
\item Epigenetics is the study of the changes in gene expression that occur without alterations in the underlying DNA sequence. Epigenetic changes are central to our understanding of key disease mechanisms including those for cancer, dementia, autoimmune disorders, along with a number of congenital deformities. As a result, significant efforts have been put in toward developing AI and machine learning methods to find patterns in epigenetic data and their relevance in regard to disease understanding. 
\item The primary goal of this article is to present a comprehensive literature review of modern AI methods for the identification and understanding of epigenetic modifications. This review is addressed to both AI experts and Epigeneticists. There are a few previously published reviews that have covered the use of machine learning in epigenetic analysis. Unlike the previous 
reviews, in this article we have approached the review process from two very different and complementary perspectives as follows:
\begin{itemize}
\item To allow the AI research community spot interesting epigenetic problems amenable to the AI methodology, we have provided a taxonomy of research problems involving 
epigenetic data that can benefit from a data-driven AI approach.
\item To provide the Epigenetics researchers with example solutions and template AI paradigms, we have mapped epigenetic problems to the class of AI models and machine 
learning paradigms that have been investigated in the literature.
\end{itemize}
\item Finally, to provide guidelines for future research, we have identified gaps in the current literature, research challenges, and possible recommendations.
\end{highlights}


\maketitle

\section{Introduction}
Epigenetics is the study of changes in gene expression, which are both meiotically and mitotically heritable modifications that occur without alterations in the underlying DNA sequence \cite{egger2004epigenetics}. The epigenetic silencing of genes is initiated and sustained by different mechanisms such as histone modifications (altering chromatin structure), DNA methylation, microRNA (miRNA) (targeting key enzymes involved in establishing epigenetic memory), and chromatin conformation \cite{skinner2014endocrine, holder2017machine, liang2018epigenetic}. Owing to these mechanisms, heritable phenotypic changes occur, which may lead to cancer, obesity, dementia, cardiac diseases, autoimmune diseases, and numerous other disorders \cite{robertson2005dna, bhusari2011insulin, soubry2013paternal}. Epigenetics is intimately related to environmental factors, making it potentially more useful for disease diagnosis and therapy than genetics alone \cite{berdasco2019clinical}. Smoking, alcohol, diet, and stress can have a significant impact on epigenetic modifications because of their role as environmental toxins \cite{joubert2012450k, anderson2012nutrition, alegria2011epigenetics}.

The primary goal of this paper is to conduct a comprehensive narrative literature review of modern artificial intelligence (AI) methods used for the identification and understanding of epigenetic modifications. This may involve variations in gene expression, changes in chromatin structure (such as nucleosome positioning), DNA methylation (such as 5-methylCpG), and histone modifications (HMs). DNA methylation changes are one of the major components of epigenetic modifications involving the addition of a methyl group to a cytosine nucleotide base which is known to play a key role in the regulation of gene expression \cite{moore2013dna}.  Methylation alters the expression by preventing the binding of transcription factors thereby restricting the transcription step leading to suppressed or no synthesis of the corresponding protein. This modulation of gene expression can lead to the progression and development of diseases such as cancer \cite{liu2020application}. The HMs are another important epigenetic mechanism that controls gene regulation \cite{alaskhar2018histone}. The presence of several histone marks along the length of the genome essentially works in a combinatorial way. Understanding these combinatorial effects is a vital step to enable the design of disease-specific interventions \cite{wang2017ethylene}. Every human cell has chromatin that stores genetic and regulatory information, where DNA in the cell nucleus is securely packed and wrapped around histone proteins. It plays a crucial role in DNA repair and replication, regulating gene expression, biological pathways, and finally complex phenotypes are all affected by chromatin structure. Therefore, DNA methylation, chromatin structure, and HMs comprise the key epigenetic mechanisms that have an essential role in the control of gene expression processes, development, and disease. 
Several recent studies have presented detailed information on the clinical potential of epigenetics. For example, the link between DNA methylation and schizophrenia is confounded by variations in smoking prevalence between patients and controls \cite{joubert2012450k}. Epigenetic mechanisms have proved to be affected by adverse early life experiences, such as starvation or smoking practiced by the mother during gestation \cite{joubert2016dna}. Furthermore, epigenetic modifications can be caused by lifestyle and environmental changes including diet, nutrition, and stress levels \cite{joubert2012450k, joubert2016dna, anderson2012nutrition}. Another important epigenetic mechanism that regulates transcriptional and post-transcriptional regulation of gene expression is non-coding RNAs, which includes long non-coding RNAs and microRNAs \cite{patil2014gene}. The regulatory roles of lncRNAs can be carried out through interactions with proteins, RNA, and DNA. Their expression is frequently condition-dependent and tissue-specific, providing for context-specific gene regulation \cite{statello2021gene}. The miRNAs are tiny non-coding RNAs that mainly control post-transcriptional regulation of gene expression \cite{patil2014gene}. In addition, epigenetic biomarkers are biological markers of epigenetic modifications including HMs, DNA methylation, and non-coding RNA expression, have emerged as a potential tool for accurate identification and diagnosis of multi-modality diseases. They are appropriate for use in clinical practice because of their easy accessibility and simple detection methods which enhance disease diagnosis, prognosis, and therapy monitoring \cite{garcia2017epigenetic, bock2008computational}. 

Over the last few decades, high-throughput experimental approaches have been used to analyze epigenetic changes. For example, Hi-C and ChIA-PET are two methods \cite{dryden2014unbiased, rao20143d} used for detecting enhancer-promoter interactions across the genome. Microarrays, RNA-seq and quantitative polymerase chain reaction (qPCR) \cite{vanguilder2008twenty} are used to identify and measure the target gene expression level. But these laboratory experimental approaches and biochemical processes are expensive and time-consuming. Due to technological advancement and the large number of annotated biological sequences, it is very difficult or sometimes impossible to identify the sequences using these conventional methods. 
AI approaches have been recently utilized to speed up the identification process in a reliable manner \cite{gupta2021artificial}. Various machine learning (ML) methods have been extensively employed in the prediction and identification of epigenetic modifications \cite{holder2017machine, rauschert2020machine}. The computational methods based on traditional machine learning have shown promise; however, they are strongly dependent on hand-designed features and require domain knowledge to extract patterns from raw data. Deep learning (DL) approaches can mitigate this limitation of hand-crafted features by learning feature representation from the data to help training classifiers \cite{lecun2015deep}. In recent years, due to the rapid development of DL algorithms, various DL methods have been developed for the extraction of useful information from epigenetic data which have shown state-of-the-art performance. In this review paper, we provide a comprehensive narrative review of recent advances in epigenetic sequence analysis with AI and deep learning approaches. 

There are a few previously published reviews that have covered the use of ML and DL models in epigenetic analysis \cite{holder2017machine, rauschert2020machine, talukder2021interpretation}. Unlike the previous reviews, in this paper, we aim to approach the review process from two different perspectives as follows. (a) We provide a taxonomy of research problems involving epigenetic data that can benefit from taking a data-driven AI approach, and (b) We map these problems to the class of AI models and deep learning paradigms that have been investigated in the literature. The above is clearly a combination of approaching the literature from two very different directions. Yet, by virtue of the topic being highly interdisciplinary, there is a need to provide a review that sketches an outline for the AI research community on how to spot interesting epigenetic problems amenable to the AI methodology. At the same time, it is also important to provide the epigenetics researchers with example solutions and template AI paradigms for problems proximal to their respective research areas. 
With this end in view and having both AI experts and epigeneticists as contributing authors of this review, we have taken the following two-pronged keyword search strategy which defines our inclusion and exclusion criteria. From the perspective of surveying different epigenetic research problems, we have included the following keywords and terms in our search queries for selecting the papers: gene expression, epigenetic gene regulation, methylation, HMs, disease markers, transcription regulation, enhancer-promoter interaction, chromatin states, and chromatin reorganization. From the perspective of reviewing the investigation of the above problems using AI/ML/DL, we have combined the above keywords with the following terms: supervised learning, deep learning, convolutional neural network (CNN), generative adversarial networks (GAN), recurrent neural network (RNN), long short-term memory (LSTM), unsupervised learning, autoencoder (AE), and transformer networks. As such, our exclusion criteria was not to include the epigenetic studies which are not based on an AI or a deep learning methodology. The review presented in this paper is based on the outcome of the above search strategy using academic databases such as Google Scholar and PubMed covering literature published till August 2024.

\begin{figure*}[!t]
\centering	\includegraphics[width=\linewidth]{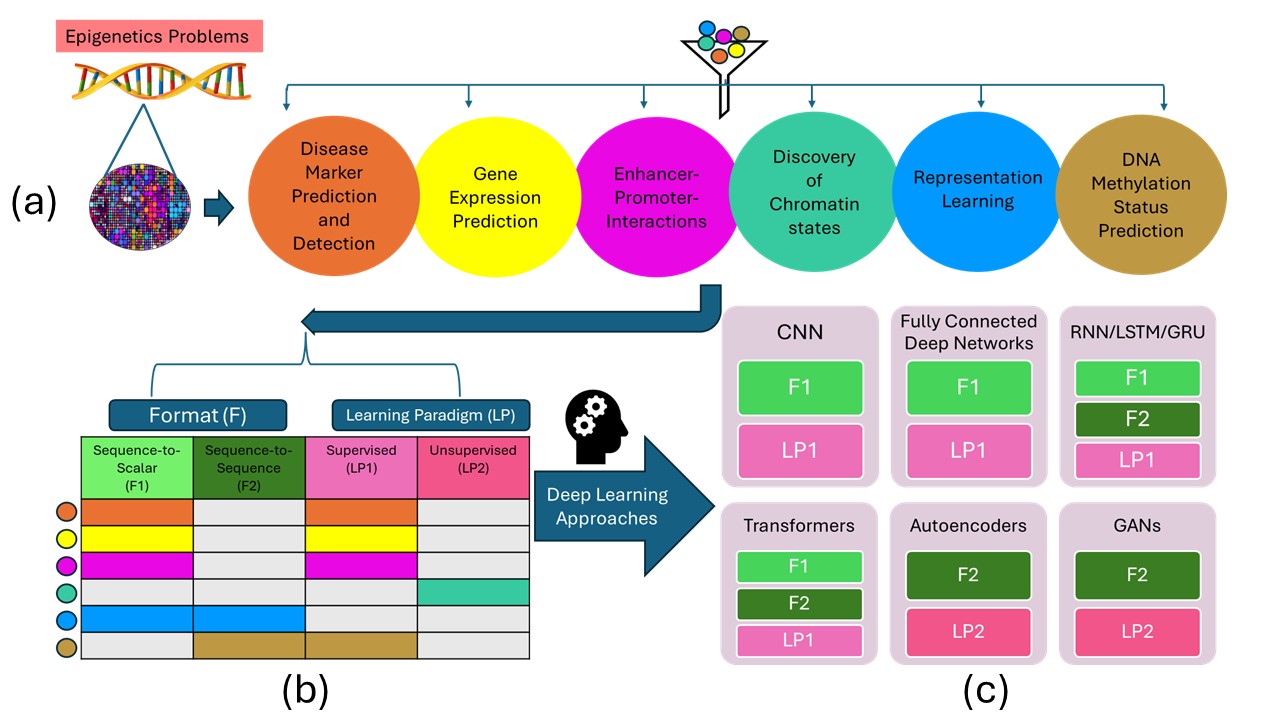}
\caption{Graphical overview of the taxonomy of various problems in epigenetic sequence analysis covered in this review article. (a) Different types of epigenetic problems. (b) A tabular illustration of which problem corresponds to which format (F) and learning paradigm (LP), e.g., first row shows that `Disease marker prediction and detection' (orange) corresponds to Format 1 (F1: Sequence-to-scalar) and Learning Paradigm 1 (LP1: Supervised). (c) Illustration of which neural network architecture corresponds to which F and LP, e.g., Autoencoders in the context of sequential inputs would correspond to F2: Sequence-to-sequence and LP2: Unsupervised.}\label{Fig1}
\end{figure*}

Figure \ref{Fig1} shows a graphical overview for the taxonomy of research problems in epigenetic sequence analysis based on the literature reviewed, wherein we mention the machine learning paradigm they fall in, along with the nature of the model’s input data and desired output depending on the research question addressed. Readers who are more familiar with epigenetics research can focus on part (b) of the figure which is a tabular illustration of correspondence between epigenetic problems and data format as well as the learning paradigm. AI researchers and practitioners may find it convenient to focus on part (c) of the figure which shows which neural network architectures correspond to which learning paradigm and format. For each of the different kinds of problems, we will review papers which have proposed specific deep learning and neural network architectures to address these problems. As a further critique and guidelines for future research, we have identified gaps in the literature, research challenges, and possible recommendations.

Our review is organized as follows. Since the choice of the deep learning architecture is determined by the nature and format of the model’s input and output, we will begin with a brief review of the nature of input data that originate in problems pertaining to epigenetics (Section \ref{sec2}). In Section \ref{sec3} we will provide a review of different deep learning methods. The next section will describe the research problems as taxonomized in Figure \ref{Fig1} along with the relevant works under each head (Section \ref{sec4}). We will conclude the review with the identification of research challenges in the field and possible future directions. 

\section{Nature of Epigenetic Data and Available Datasets}\label{sec2}
In the previous decades, microarray was one of the most popular sequencing techniques for acquiring large amounts of data on gene expression patterns throughout the whole genome \cite{tao2017microarray}. Affymetrix \cite{gohlmann2009gene} and Illumina \cite{barnes2005experimental} are the two most powerful microarray platforms. However, there are many prominent microarray producers, including Taqman \cite{taqman_website}, Exiqon \cite{exiqon_website}, and Agilent \cite{zahurak2007pre}. Currently, large microarray gene expression databases are available online at various public repositories and microarrays that enable the simultaneous analysis and measurement of the expression of a large number of genes \cite{moore2013dna,castillo2017integration,bernstein2007mammalian,aryee2014minfi}. The methylation of the cytosine nucleotide base in CpG islands is one of the key epigenetic factors affecting the expression of genes \cite{aryee2014minfi}. Illumina Human Methylation Infinium Bead Array is a widely used technique to measure and determine the DNA methylation status in the whole genome \cite{kurdyukov2016dna,triche2013low}. The Illumina technology \cite{kurdyukov2016dna} is cost efficient and allows scanning a bigger part of the genome; in particular, the number of CpG sites used to range from 27000 to 450000, which recently has been increased to 850000 with the EPIC array \cite{bibikova2009genome,sandoval2011validation,moran2016validation}.  From a data format perspective, the methylation status of CpG sites can be considered as a vector or array of numbers. Similarly, further methods in the field of epigenetic modifications are bisulfite sequencing data analysis and chromatin immunoprecipitation followed by sequencing (Chip-seq) \cite{wreczycka2017strategies,krueger2012dna,xu2010application}. Bisulfite sequencing is a technique widely used in epigenetics research to accurately examine and determine the DNA methylation patterns in various contexts such as disease state and employs sodium bisulfite for converting unmethylated cytosines to uracil, while the original methylated cytosines are unchanged \cite{wreczycka2017strategies}. RRBS (reduced representation bisulfite sequencing) and WGBS (whole-genome bisulfite sequencing) are the two most comprehensive bisulfite sequencing methods used for the investigation of methylation data \cite{sun2014moabs}.

In addition, ChIP-seq is a powerful technique used for mapping HMs, transcription factors (TF), chromatin regulators, histone proteins, and other DNA-binding proteins. It has contributed significantly to our knowledge of disease processes and the examination of epigenetic modifications for prospective clinical applications. The data resulting from Chip-Seq (e.g. HMs) can be considered as a multi-channel sequential data aligned with base-pair indices. There are several public databases namely ENCODE \cite{encode2012integrated}, ROADMAP epigenome database \cite{kundaje2015integrative,bernstein2010nih}, epigenome database for human endothelial cells \cite{endothelial_database}, and Chip-Atlas \cite{oki2018ch} which can lead to various data formats including raw read file (FastQ), mapped read file (BAM), peak files, quality check results, and gene expression data. These files contain both RNA-seq and Chip-seq data. These are the most popular and publicly available genomic and epigenomic databases, omics resources, and repositories which provide comprehensive information and resources about the genome and epigenetics to the researchers. For instance, the ROADMAP epigenome database contains high-quality, genome-wide maps of several key HMs, DNA methylation, mRNA expression, and chromatin accessibility across hundreds of human cell types and tissues, and overall has data spanning 150.21 billion map sequencing reads \cite{kundaje2015integrative}. Researchers can access and analyze these databases or subsets of data programmatically through Application Programming Interfaces (APIs), which allows them to import only the relevant data into their programs or scripts without fully downloading it. For example, the ENCODE database provides a complete ENCODE REST API \cite{encode_api} that allows researchers to retrieve genomic data such as epigenetic modifications, TF binding sites, and expected genome area. 

\section{Deep learning }\label{sec3}
\begin{figure}[h]
\centering	\includegraphics[width=\linewidth]{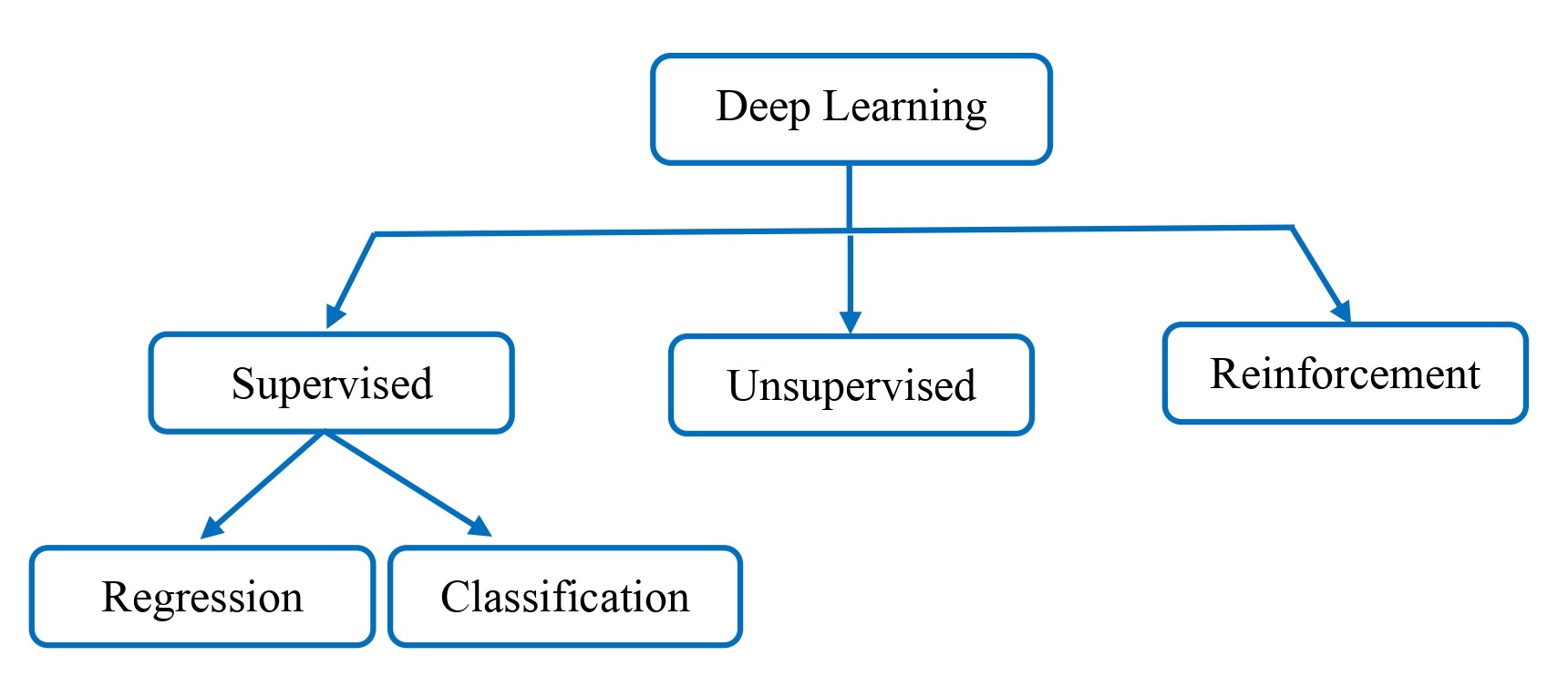}
\caption{Different learning paradigms for DL: supervised, Unsupervised, and Reinforcement learning}\label{Fig2}
\end{figure}

Deep learning is a specialized branch of machine learning that can automatically extract features from raw input data, without human-engineered features \cite{lecun2015deep}. It relies on deep neural network (DNN) architectures, which have been demonstrated to have state-of-the-art performance in diverse domains such as image processing \cite{krizhevsky2012imagenet}, natural language processing \cite{sutskever2014sequence}, and speech recognition \cite{abdel2014convolutional}. DL holds significant promise in bioinformatics, facilitating the analysis of large-scale high-throughput epigenomic data \cite{xiong2015human, kundaje2015integrative}, predicting gene expression \cite{singh2016deepchrome, zhang2022transformer}, disease classification \cite{li2021dismir}, and enhancer-promoter interactions \cite{mao2017modeling,singh2019predicting}. DL algorithms are broadly categorized into three types (Figure \ref{Fig2}): supervised learning, unsupervised learning, and reinforcement learning \cite{zhang2020survey,kiran2018overview,jaiswal2020survey}. 
In supervised learning, labelled data are available and the models are trained to minimize the discrepancy between a model’s predictions and desired outcome (as determined by the ground-truth). This setup is usually used to solve classification and regression problems. Unsupervised methods are invoked when annotated data are not available. In the absence of labelled data, the data can still be grouped into clusters \cite{elhassani2022deep,mantach2022deep, zou2020sequence}, and useful representations can be learned through autoencoders \cite{norouzi2024volpam}. In addition to unsupervised representation learning, recently self-supervised methods such as contrastive learning have been used for learning representations \cite{khosla2020supervised,xie2022self, liu2023self}. The learned representations through unsupervised methods can then be deployed later for training supervised models if annotations are available for a smaller subset of data \cite{yakimovich2021labels}. In reinforcement learning, learning and collection of training data happens concurrently while an agent interacts with the environment, collects data, and receives feedback, which in turn is used to train model parameters to maximize a reward function \cite{sutton2018reinforcement}. Moreover, recent years have seen the emergence of generative models which fall within the category of self-supervised methods, and are trained to generate data that adhere to a training data distribution. Common generative models include Generative Adversarial Networks (GANs) \cite{goodfellow2020generative}, Variational Autoencoders (VAEs) \cite{kingma2013auto}, and diffusion models \cite{croitoru2023diffusion, kingma2021variational}.

One major factor behind the choice of a neural network architecture is the nature of input data. For example, if the input consists of feature vectors, fully connected networks (FCNs) can be employed \cite{lin2015far}. For images, the go-to choice would be convolutional neural networks (CNNs) \cite{sultana2020evolution,o2015introduction}. If the input contains sequential data, the recommended architectures would include Transformers \cite{vaswani2017attention}, Recurrent Neural Networks (RNN) \cite{lecun2015deep}, and Long Short Term Memory (LSTM) models \cite{hochreiter1997long,lecun2015deep}. The other consideration is the choice of the learning paradigm e.g., supervised, unsupervised, or semi-supervised depending on the problem and the availability of labeled data \cite{mantach2022deep}.
Since data originating in epigenetics research problems can have a variety of forms, we briefly introduce some  of the common neural networks as below.

\subsection{Convolutional Neural Networks}\label{subsec1}
\begin{figure*}[!t]
\centering	\includegraphics[width=\linewidth]{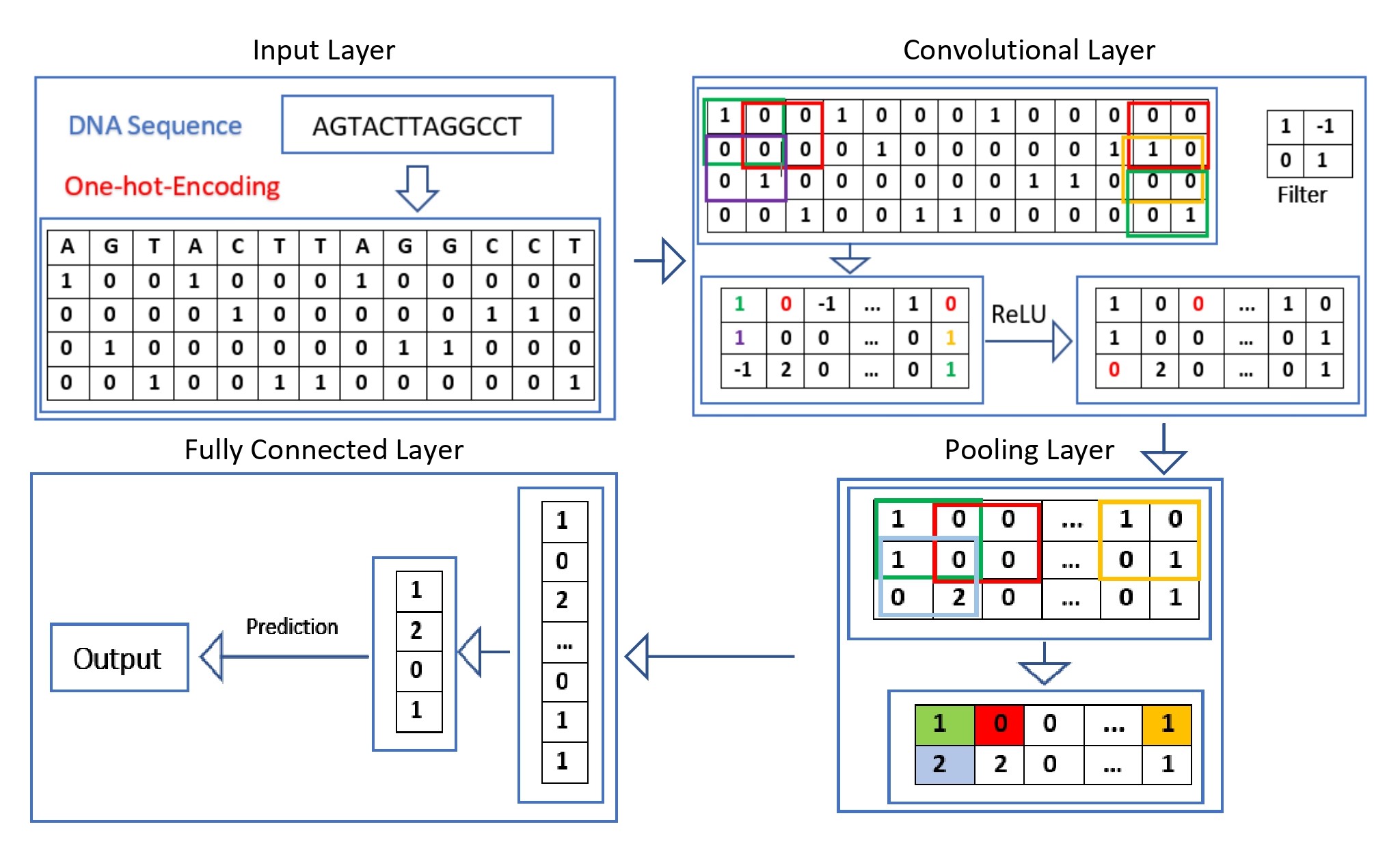}
\caption{An example of CNN consists of the input (DNA sequence) used one-hot-encoding, a convolutional, pooling, and fully connected layers with output.}\label{Fig3}
\end{figure*}

CNNs constitute a prominent class of DL neural network architecture which have proven to be highly effective in various fields such as computer vision, image processing, natural language processing, and bioinformatics \cite{bai2023mlacnn,roth2015improving,zhang2019deep, wang2022deepac4c}. While CNNs were initially proposed for imaging applications \cite{alzubaidi2021review}, and hence 2D or 3D CNNs are better known \cite{lecun2015deep,yamashita2018convolutional,khoshdel2020full}, we will review a 1-dimensional (1D) CNN which is more prevalent in genomics \cite{lv2021convolutional}. In particular, a multi-channel 1D CNN has become a widely employed DL architecture for sequential data involved in genomics research, effectively applied to analyze sequence data and decode genomic and epigenomic patterns. The CNN architecture is characterized by its unique layer structure, beginning with an input layer, followed by a convolutional layer responsible for generating feature maps. The convolution operation essentially involves computing a weighted sum over local neighborhood of the input as weighted by a set of learnable filter parameters. Typically, a convolutional layer involves passing the input through multiple filters, and the filter responses are referred to as feature maps. Subsequently, a pooling layer is used to reduce the spatial dimensions and retain significant information. Usually, a CNN constitutes a cascade of convolutional and pooling layers, eventually culminating in a fully connected layer employed to make predictions or decisions based on the extracted features, as shown in Figure \ref{Fig3}. Overall, the goal is to optimize the parameters of all the layers such that a loss function based on the output of the final layer is minimized. Examples for loss function include classification loss such as cross-entropy or regression loss such as mean-squared error \cite{khoshdel2020full,zhuang2019simple}.

To be used as an input to a CNN, it is customary to convert a nucleotide sequence into a 4x1 1-hot representation, wherein a 1 represents the particular nucleotide (Figure \ref{Fig3}). Other aspects, such as HMs can be added as additional channels. Despite handling multiple pieces of information, such a network is still 1D CNN because inherently there is only one dimension or coordinate, which is the base pair index, while the information in different channels are attributes associated with the same base pair index.

\subsection{Recurrent Neural Networks}\label{subsec2}
\begin{figure}[!h]
\centering	\includegraphics[width=\linewidth]{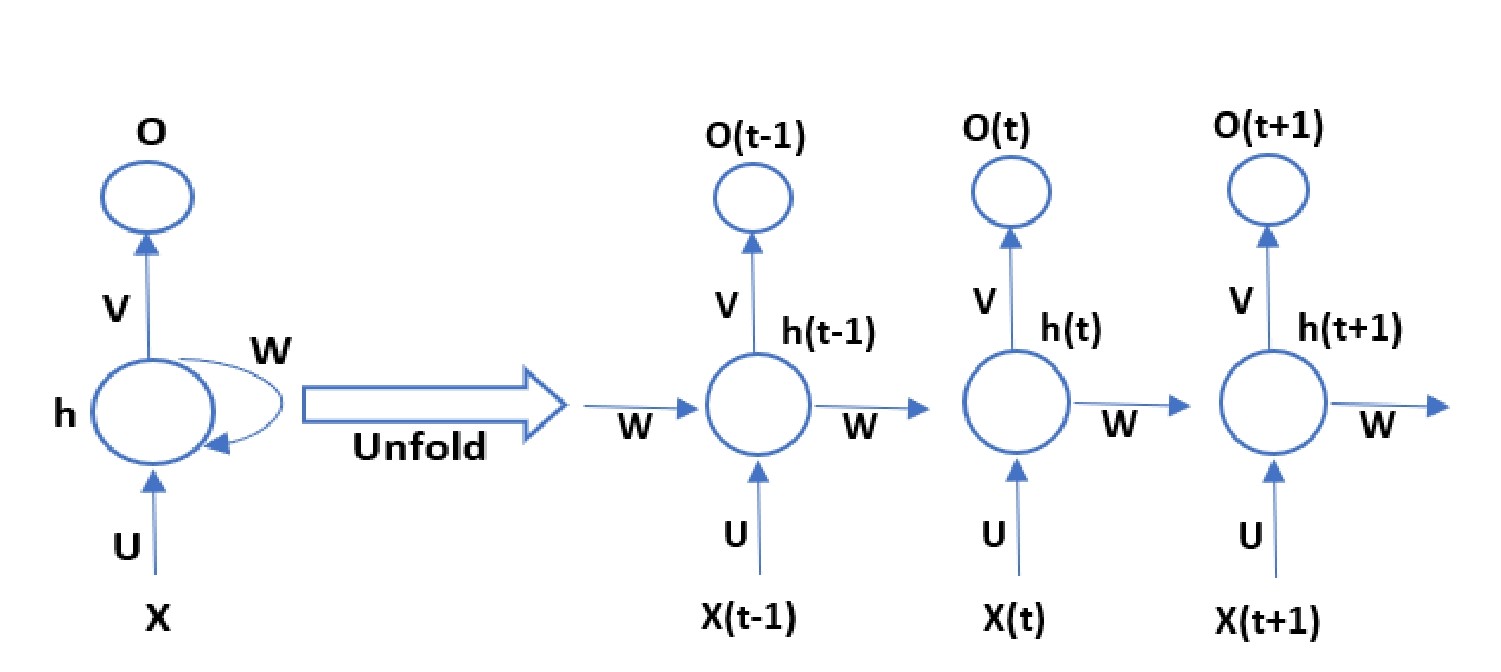}
\caption{A representation of the RNN architecture with its corresponding functional components such that X is the input layer, h is the hidden layer (h(t) and h(t-1) are new and previous states), O is the output layer. U, V, and W represent the model parameters.}\label{Fig4}
\end{figure}

The RNN architecture is specifically designed for sequential data processing, such as text and genomics, and is capable of preserving state information across time steps \cite{lecun2015deep} or a variable that marks the coordinates of a sequence such as base pair index. In this architecture, the output of a previous state serves as input to the current state, enabling the network to depend on past information while learning the current context. Comprising three main layers, the RNN architecture consists of an input layer, a hidden layer with recurrent connections, and an output layer \cite{elhassani2022deep}. The input layer receives sequential data as input, and the hidden layer processes these data while retaining information from previous time steps. In RNN, the output layer utilizes sequential information to generate the desired prediction output, as shown in Figure \ref{Fig4}. The figure portrays the functional components of the RNN structure, highlighting the analysis of sequential data in time domain for comprehending its temporal dependencies if the sequence is indexed by a time variable, or spatial dependencies if the sequence is indexed by a spatial coordinate, thereby supporting several application domains including signal analysis, natural language processing, and genomic analysis \cite{chen2024multi,wang2024birnn}. 

In RNN the role of hidden states is pivotal due to their capability of capturing and persevering time dependencies. However, this preservation is contingent on the length of the input history since for longer sequences the model suffers from the problem of vanishing gradients \cite{Goodfellow-et-al-2016}. Vanishing gradients occur due to the model weights assuming small values resulting in the shrinking of gradient values with each successive computational step.
The \textbf{Long Short-Term Memory (LSTM)} architecture is a widely used model for sequential data that can handle the vanishing gradient issues more efficiently as compared to simple RNNs based on conventional activation functions such as sigmoid and tanh \cite{lecun2015deep,hochreiter1997long}. The LSTM employs the concepts of storage units and controlling gates to handle the long-term dependencies across the data processing pipeline during the model learning process. In these models, the storage units and the gates are based on the neurons such that each of the neurons holds a storage unit and three data placeholders known as input, forget, and output gates, which learn the relative importance of data over time. In the network, these gates regulate the flow of information in such a way that helps to avoid the vanishing gradient problem. The input, forget, and output gates perform the role of regulating the information flow within the memory cells, the decision whether to retain or discard information, and control of information outflow in conjunction with monitoring the flow of information within the layers. The LSTM architecture's capability to handle long-term dependencies makes it a powerful tool for sequential data analysis, leading to its widespread applications in diverse fields, including time series forecasting, natural language processing, and speech recognition and bioinformatics \cite{ubal2023predicting,yin2017comparative, hamdy2023deepepi}. Other variants of RNNs also exist such as Gated Recurrent Units (GRU) \cite{chung2014empirical}, and bidirectional LSTM/GRU wherein a sequence is traversed in both directions and hidden states are maintained for forward and backward traversal allowing to discover more rich sequential patterns \cite{canatalay2022bidirectional,li2022identifying, chen2022deepm6aseq}. 

\subsection{Autoencoders}\label{subsec3}
\begin{figure}[!h]
\centering	\includegraphics[width=\linewidth]{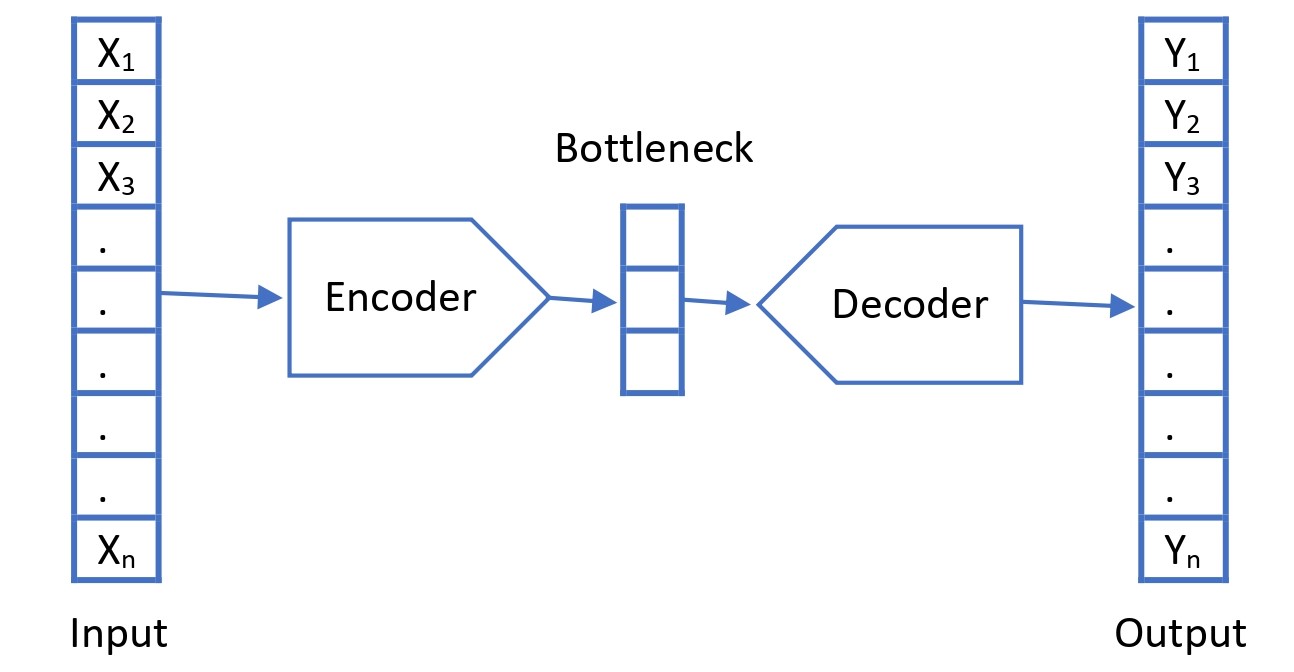}
\caption{Architecture of an autoencoder consists of an input layer containing $n$ elements from $X_{1}$ to $X_{n}$, encoder, bottleneck, decoder, and output layers containing $n$ elements from $Y_{1}$ to $Y_{n}$ which are reconstructed from original input data.}\label{Fig5}
\end{figure}

Autoencoders (AE) represent an unsupervised learning technique used for representation learning \cite{mantach2022deep,kingma2013auto}. The architecture of AE consists of three main layers: the encoder, bottleneck, and decoder. The encoder layer aims to change the dimensionality of input data and convert it into a latent representation \cite{shi2022toxmva}. The bottleneck layer holds the compressed information of the input data, while the decoder layer reconstructs the original input data from the latent representation. Usually, the hidden layer (bottleneck) has a reduced number of neurons compared to the input and output layers. The layers before the bottleneck serve as the encoding function, while the layers after it function as the decoding part. Training of autoencoders involves employing the backpropagation approach to minimize the model's loss, which is calculated as the difference between the input and output. A visual representation of the AE architecture can be seen in Figure \ref{Fig5}.

Once trained, the output of the encoder can be used as a feature representation and is typically employed for supervised learning tasks \cite{suryawati2021unsupervised}.

\subsection{Transformers}\label{subsec4}
\begin{figure*}[!t]
\centering	\includegraphics[width=\linewidth]{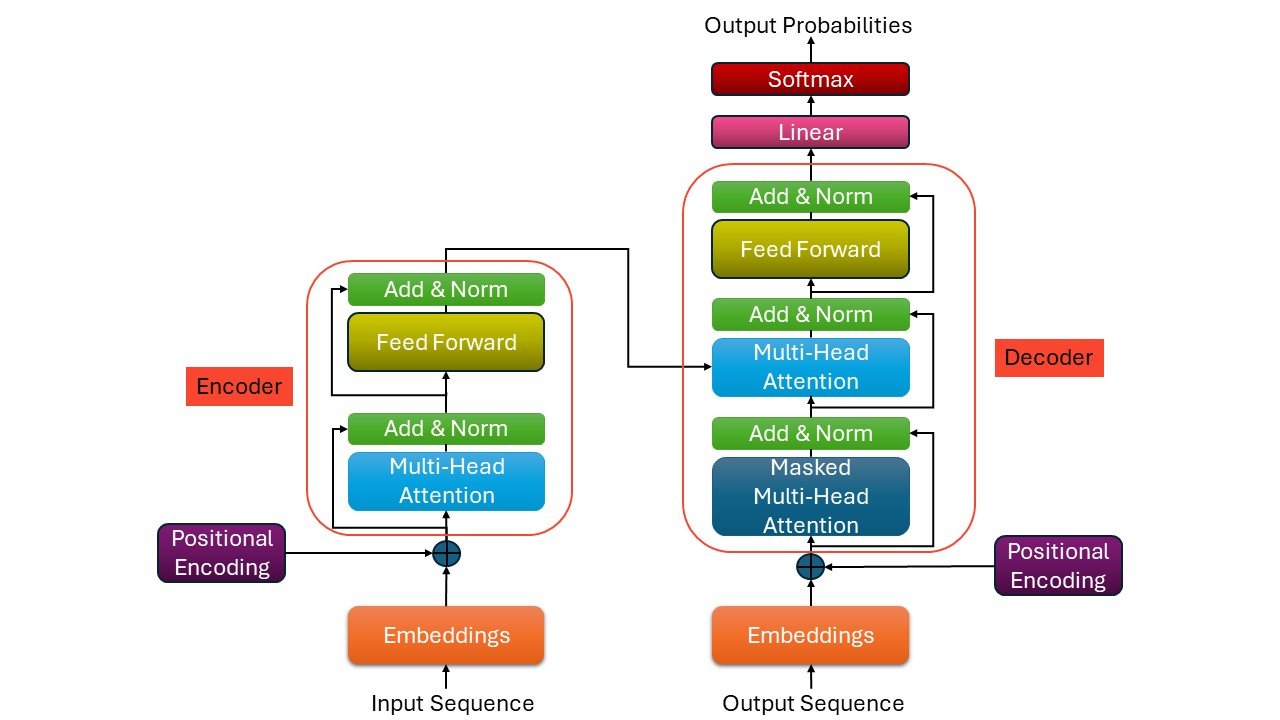}
\caption{An architecture of Transformer consists of the input sequence, encoder, decoder and output.}\label{Transformer}
\end{figure*}

The transformer is a cutting-edge deep learning architecture initially designed for processing textual data and has shown exceptional performance in various NLP tasks, such as machine translation \cite{vaswani2017attention}. The transformer architecture comprises two main components: the encoder, and the decoder, each consisting of multiple layers of self-attention and feed-forward neural networks, as shown in Figure \ref{Transformer}. In RNNs, the computations are by necessity serial, since a hidden state at a particular time cannot be updated unless it has received the hidden state from the previous time point, and the input from the current time point. Transformers, instead allow parallel computations over the entire sequence through positional encoding and self-attention. The self-attention mechanism enables the model to focus on and assign scores to relevant data points within a given context. The feed-forward neural network then applies non-linear transformations to the outputs of the self-attention mechanism. The encoder processes the input sequence to generate a set of hidden representations, while the decoder utilizes these hidden representations to produce the output sequence. The information is encoded by the stacked encoders and decoded by the stacked decoders, with the stack size determined by the architectural design. In addition to its success in NLP tasks such as machine translation, the transformer model has been effectively applied to enhance prediction performance in various domains, including genomics, protein-to-protein interaction, and others \cite{ahmed2024epi, wu2024attentionmgt, pei2023identification, Tahir2024TransformerChrome}.

\section{DL methods for epigenetic problems}\label{sec4}
In this section, we develop a taxonomy of various problems in epigenetic sequence analysis reported in Figure \ref{Fig1}, map them to deep learning methods as required by the nature of the data and the question addressed, and then review papers under each head below:

\subsection{Disease Marker Prediction and Detection}\label{subsec5}
The DNA methylation states are known to be altered throughout the genome in the early stages of a cancer, which allows the use of methylation status as a valuable feature for early detection of cancers \cite{kulis2010dna}. As such, methylation states at distinct CpG sites, or in various sub-genomic regions, can be combined to form a feature representation to build and train improved cancer detection models. In this regard, Li et al. \cite{li2021dismir} designed a DL-based method called DISMIR (Deep Integrating Sequence Individual Reads) for ultrasensitive detection of cancer. DISMIR method uses methylation information and DNA sequence with plasma cell-free DNAs WGBS data. In this method, a novel feature representation called switching reads and switching regions was introduced to discover cancer-specific differentially methylated regions (DMRs), that improve the read-resolution of cancer-related signals. Switching regions and switching reads were used to identify cancer-specific DMRs across the entire genome. The DISMIR model uses both the genomic sequence as well as the corresponding methylation status represented as the one-hot encoding for nucleotide bases and methylation. This representation is passed through a bi-directional LSTM followed by a 1D CNN, and produced  a real value between 0 and 1 which represents the probability of disease. DISMIR has the potential to be an accurate and reliable non-invasive early-stage method for different types of cancer detection and obtained mean AUC ROC of 0.9969 and 0.9112. 

Liu et al. \cite{liu2019dna} established a DL-based predictive method using CpG methylation markers data of 27 diverse cancer types collected from The Cancer Genome Atlas (TCGA) \cite{weinstein2013cancer} and Gene Expression Omnibus (GEO) \cite{barrett2012ncbi} datasets. In this model, the authors employed a t-statistic based approach to collect the top 2000 CpG sites as candidate markers from 485,000 original CpG sites. The candidate markers consist of 2000 promoter markers and 2000 CpG markers. Then, LASSO (least absolute shrinkage and selection operator) and random forest algorithm \cite{tibshirani1996regression, hassan2023comparative} were employed to further refine the list of candidate markers to 13 promoter and 12 CpG markers. These final markers were employed to train two multi-layer feedforward neural network models. This model obtained AUC ROCs of 0.995 and 0.993 for promoter and CpG markers, respectively to predict pan-cancer accurately. 

Albaradei et al. \cite{albaradei2021metacancer} developed a DL-based method namely: MetaCancer to predict pan-cancer metastasis status. This model used three heterogeneous data types from TCGA containing DNA methylation data, RNA sequencing, and microRNA sequencing from 400 patients. The MetaCancer method automatically extracted features by convolutional variational autoencoder and then employed a deep fully connected network to identify tumours as primary or metastasized. The result showed that the MetaCancer method significantly outperformed the existing SVM ensemble method on various metrics (accuracy of 88.85\% versus 82.50\%). Zhang et al. \cite{zhang2021omiembed} designed a DL-based method namely: OmiEmbed to predict cancer survival. In addition, this model also enabled multitask learning such as multi-omic integration and dimensionality reduction, clinical and demographic feature reconstruction, tumour type classification, and survival prediction. The OmiEmbed model contains two modules i.e., a deep embedding module and a downstream task module. The deep embedding module used a variational autoencoder (VAE) to transform multi-omic data with high dimensionality into a low dimensional latent space and fed it into a downstream task module. Then, in the downstream task module, a multi-layer fully connected network was trained using the latent representation (i.e. encoder's output) to predict the primary site and disease stage as well as classify the tumour type. The result showed that OmiEmbed method outperformed other machine learning methods (AUC ROC of 0.9943 versus 0.9863). 

Xiao et al. \cite{xiao2021cancer} developed a DL model based on the Wasserstein generative adversarial network (WGAN) to generate data and predict cancer cases from imbalanced datasets because it is a common issue in diagnostic application. This method predicted the gene expression data of breast, lung tissues, and stomach. The result showed that the proposed model improved the predictive performance on all three datasets as compared to previous methods for imbalanced data, such as random oversampling, SMOTE \cite{chawla2002smote, li2024metaac4c} technique (accuracies: 98.33\%, 96.67\%, and 96.67\%).  Manzanarez-Ozuna et al. \cite{manzanarez2018model} designed a DNN-based model to predict mRNA-Smad7 expression regulation by miRNAs using the expression values of 179 mRNA-Smad7 and miRNAs in 1074 samples of breast cancer patients. A genetic algorithm (GA) was employed to find the optimal design for a deep neural network model with efficient predictive performance, as well as to find or select features. The authors selected 44 miRNA sequences to train their model. Then, the Olden algorithm \cite{olden2002illuminating,olden2004accurate} was used to determine the relative relevance of each of these miRNAs on the expression of mRNA-Smad7. To evaluate the importance of features, the Olden algorithm for assessing variable significance was applied \cite{manzanarez2018model}. Specifically, the algorithm makes use of all the connection weights in a deep neural network considering both the direction and the magnitude of the signal's excitation \cite{manzanarez2018model}. The DNN identified 23 miRNAs that contributed the most in its predictions, in which five have been experimentally verified to be connected with breast cancer. Rajpal et at. \cite{rajpal2023xai} developed an AI-based method, XAI-MethylMarker, to discover biomarkers for breast cancer subtype classification based on methylation data. This method involves a two-stage framework to discover 52 distinct differential DNA methylation biomarkers for the classification of breast cancer subtypes using a feed-forward neural network and an autoencoder in a DL network. The result showed that the XAI-MethylMarker method significantly outperformed the existing random forest and bootstrapping ensemble method (accuracy of 0.8145 versus 0.7530). 

In a recent study, DeepHistone \cite{yin2019deephistone} computational model used chromatin-accessible signal and DNA sequences to predict histone modifications sites. The DeepHistone model consisted of three modules: DNase module (chromosome accessibility module), DNA module (sequence module), and a joint module. The sequence module used one-hot encoding to convert DNA sequence of length (L) into L×4 matrix and fed it as input matrix to a 1D CNN model to extract hidden features. Similarly, chromosome accessibility module used the same CNN model architecture of DNA modules to extract features. Finally, the feature space obtained from these two modules was fed to the joint module for classification. In addition, the DeepHistone model has been reported to identify biologically important motifs and functional motifs in the cell lines studied. Several patterns discovered in different cancer cell lines correspond to motifs that have been previously linked to specific cancer types. For example, DeepHistone retrieved the E2F3, a transcription factor identified to be overexpressed in lung cancer tissue, from a lung cancer cell line \cite{yin2019deephistone}. Furthermore, in a cervical cancer cell line, the DeepHistone model also found that NR2F6 and PROX1 were highly correlated with the progression and spread of cervical cancer. The paper reported that the DeepHistone model outperformed previous methods (average AUC ROC of 0.9065). 

Similarly, Baisya et al. \cite{baisya2020prediction} developed a DL model, DeepPTM, to predict histone Post-Transcription Modification (PTM) from DNA sequences and TF-binding data. DeepPTM employed a feed-forward fully connected neural network on TF-binding Chip-Seq dataset and DNA sequence datasets, respectively.   The aforementioned neural network was trained to produce a probability for histone PTM as an output. The DeepPTM used four histone markers rather than seven used by DeepHistone and showed better AUC ROC performance than DeepHistone (average AUC ROC of 0.9543 versus 0.9132 on four histone markers). Zhang et al. \cite{zhang2022transformer}, introduced a DL method namely T-GEM (Transformer-Gene Expression Modeling) for immune cell type classification and cancer type prediction using a transformer neural network. The T-GEM model used multi-head self-attention modules to identify and capture the most important biomarkers across various cancer subtypes to handle the complexity of high-dimensional gene expression.  
\input{Table1}

Jiang et al. \cite{jiang2020generative}, developed a DL model for disease gene prediction called GAN-DAEMLP (Generative Adversarial Network- De-noising Auto-encoder and MultiLayer Perceptron) using mouse RNA-seq data.   In their model, they coupled generative adversarial network (GAN) and de-noising auto-encoder (DAE) such that the GAN was utilized as a generator and MLP was employed as a discriminator. The GAN-DAEMLP was able to differentiate between healthy and disease samples while also providing a risk score. Their experimental results showed the superiority of the GAN-DAEMLP by identifying ten different types of disease-related genes. The result showed that the GAN-DAEMLP model improved the prediction performance with AUC ROC of 0.6700. Most recently, a deep learning-based model called iHMnBS (identification of HMs and Binding Sites) was designed to determine which of the seven HMs a DNA sequence may bind with, as well as which portions of the DNA sequence bind to them \cite{li2022identifying}. This model contained DNA processing and DNase processing modules, used to process two types of input data such as DNA sequences and TF-binding Chip-Seq data. The model then employed a CNN (DenseNet) to automatically learn hidden features from the DNA sequences and DNase-seq, respectively, and concatenated these two feature spaces.  Based on the results, iHMnBS model outperformed the existing DeepHistone model (average AUC ROC of 0.9411 versus 0.9065). The performance comparison of the above stated models is shown in Table \ref{tab1}. As can be noted in the table, for some of the methods, exceptionally high values have been reported for performance metrics e.g., a perfect AUC of $0.99$. While the techniques used in these works are sound, there is need to revisit the diversity of the datasets on which the results have been reported. In addition, to establish the robustness of the results more stringent cross-validation strategies should be explored. We will address these issues in detail in Section \ref{sec5} (Challenges and Recommendations).

\subsection{Gene Expression Prediction}\label{subsec6}
The gene expression of a particular gene is determined by the amount as well as the synthesis rate of its downstream functional product such as protein or RNA \cite{liu2020fully}. The process involves generating a functional RNA using genetic information from genes and determining what regions of the genome are transcribed. Many factors at different levels influence gene expression such as variations in the non-coding part of DNA, methylation status, and HMs etc \cite{liu2020application}. In this section, we discuss models based on DL for prediction of gene expression using varied inputs depending on the application context. In this regard, Singh et al.\cite{singh2016deepchrome}, developed DeepChrome, the first deep convolutional neural network-based discriminative framework to predict gene expression and also attempt to interpret the epigenetic factors involved in gene regulation.  DeepChrome's primary objective was to determine gene expression using five HMs marks from several human cell types. Specifically, gene expression levels and five types of HMs (H3K27me3, H3K9me3, H3K36me3, H3K4me3, and H3K4me1) signals were used to train the DeepChrome model from 56 various types of a cell based on data derived from the REMC database \cite{kundaje2015integrative}. The input to the model was a binned version of histone marks wherein the binning was done over a 10,000 base pairs region centered around the transcription start site (TSS) of each gene into bins of 100 base pair length. Specifically, a region spanning \(\pm5000\) basepairs on both sides of the TSS was binned into 100 bins each consisting of average values of histone marks from 100 base pairs. The DeepChrome model consistently outperformed previous methods such as random forests \cite{dong2012modeling} and SVM \cite{cheng2011statistical} (average AUC ROC of 0.8008 versus 0.59 and 0.66) on 56 cell types. 

The research group that developed the DeepChrome model also later proposed another model namely, AttentiveChrome \cite{singh2017attend}, utilizing the same benchmark datasets of 56 cell types. Hierarchical attention-based LSTM models were used in AttentiveChrome to investigate dependencies between chromatin factors that controlled gene regulation. Here, after analyzing the five core histone marks, the H3K36me3 mark was considered as gene body structure, H3K4me3 and H3K4me1 were considered as promoter and enhancer marks, and the H3K9me3 and H3K27me3 were defined as the repressed gene markers. During the training, the AttentiveChrome method employed the two levels of soft attention mechanism \cite{bahdanau2014neural} i.e., one for essential chromatin markers and the other for significant spots within those markers to predict gene expression. Using these attention layers, the attention weights provided insight into the portions of the input that the model relied on the most for making classification decisions. The attention weights for expressed genes were found to be high corresponding to the enhancer, gene structure markers, and promoter, while average or low valued attention weights were observed around the represser markers. The genes that were not expressed displayed the opposite result. Finally, the AUCROC for the AttentiveChrome method was better as compared to the DeepChrome method on 50 cell types out of 56. Moreover, the average AUCROC for AttentiveChrome was 0.8133 as compared to 0.8008 for DeepChrome. 

Another model for predicting gene expression from histone marks was DeepDiff \cite{sekhon2018deepdiff}. Similar to the AttentiveChrome technique, the DeepDiff method was trained on the same benchmark datasets used by Singh et al. \cite{singh2016deepchrome} and employed a hierarchy of LSTMs with two levels of attention weights that were simultaneously learned. The DeepDiff used a siamese contrastive loss and multitask learning to improve the performance \cite{caruana1997multitask}. The multitask learning framework constrains the network to learn effective joint representations based on auxiliary tasks and to produce multiple predictions per sample of input data. It was first trained to identify the cell type in the sample, then used the siamese contrastive loss to enhance the learned representations.  The learned attention weights were noted for the top five predicted up/down-regulated genes in cancer cells, which corresponded to the five HM marks. The H3K4me3 and H3K4me1 histone marks obtained significantly higher weights in the up-regulated genes, while their weights in the down-regulated genes were comparatively low. In contrast, as shown experimentally in certain cell lines H3K27me3 had a higher weight in downregulated genes and a low weight in upregulated genes \cite{gregoire2016transposable}. DeepDiff produced better results as compared to AttentiveChrome in terms of gene expression prediction. 

Recently, Cheng et al. \cite{cheng2020simplechrome}, developed two models namely, DeepNeighbors and SimpleChrome, to predict gene expression. These two models were trained on the same datasets used by Singh et al. \cite{singh2016deepchrome}. The training of the DeepNeighbors model consists of two phases. First, they employed unsupervised learning i.e., Variational Autoencoders (VAEs) \cite{bunrit2020improving} to convert input matrices of each gene histone modification into lower dimensions. In phase two, the representations for both the neighboring and target genes were merged and fed into a multilayer perceptron (MLP) model to predict gene expression. The SimpleChorme model only used the first training phase of the DeepNeighbors model and excluded the neighboring genes to predict gene expression. In this study, the authors randomly selected 3 cell lines out of 56 and considered a very small size of training sample i.e., 1000 or 100 genes out of 6601 available in the dataset. Based on the results, the SimpleChrome model was shown to give performance nearly equivalent to the DeepChrome method (average AUC ROC of 0.809 versus 0.803) and lower time complexity than DeepChrome (10 sec versus 60 sec). Kamal et al. \cite{kamal2020gene} developed a model based on stacked temporal convolution networks to predict gene expression from HMs. This model transforms HMs data into a one-dimensional space and utilizes temporal convolution networks to predict gene expression.  It outperforms other models in terms of AUC ROC, recall, specificity, precision, F-Score, and accuracy (ACC).

In another study, a model called ShallowChrome \cite{frasca2022accurate} was proposed that employed feature extraction and logistic regression classifier to predict gene expression. Further, all benchmark datasets of 56 different cell types related to gene expression quantification and five types of HM marks (H3K27me3, H3K9me3, H3K36me3, H3K4me3, and H3K4me1) were extracted from the REMC database \cite{kundaje2015integrative}. In this work, the authors do not use binning approaches as used in Deepchrome and AttentiveChrome. They mentioned that the main limitation of binning is that the most predictive information may be hidden in locations that dynamically depend on the particular HM marks. The ShallowChrome method was shown to outperform methods such as DeepChrome and AttentiveChrome (average AUC ROC of 0.8737 versus 0.8008 and 0.8133, respectively) on 56 cell types.

Hamdy et al. \cite{hamdy2022convchrome} developed a deep learning-based predictive model, ConvChrome, to identify gene expression from histone modification data using REMC database. The architecture of this model consists of three main parts including 1D-CNN, 2D-CNN, and 1D-CNN followed by a self attention mechanism. The result showed that ConvChrome produced better performance than DeepChrome and Attentive chrome (average AUC ROC of 0.8399 versus 0.8008 and 0.8133). Similarly, Chen et al. \cite{chen2022predicting} developed a predictive model namely, TransferChrome, to predict gene expression from histone modifications using REMC database. This model used CNN model with self-attention mechanisms to capture global contextual information in the HMs and employed transfer learning to enhance the prediction performance for all cell lines gene expression prediction. The TransferChrome model was shown to outperform the previous ConvChrome model (average AUC ROC of 0.8479 verse 0.8399). Hamdy et al. \cite{hamdy2023deepepi} proposed a predictive model, DeepEpi, for HM based prediction of  gene expression using REMC database. This model used a CNN to detect patterns in histone signals, LSTM to capture the temporal dependencies in HMs, and then merged LSTM and CNN using ConvLSTM with a self-attention mechanism to predict gene expression. This predictive method main objective is to model complex dependencies between histone reads and long-range spatial genomic data. The DeepEpi model was shown to outperform the previous TransferChrome model (average AUC ROC of 0.8887 versus 0.8479). 

Similarly, Tahir et al. \cite{Tahir2024TransformerChrome} developed a predictive model, TransformerChrome, to predict gene expression from histone modifications. This model used transformer architecture with multi-head attention for HM marks to learn attention and feature representation for predicting gene expression. 
The TransformerChrome model was reported to outperform the DeepChrome in 39 out of the 56 cell types analyzed. Across these 39 cell types, the TransformerChrome model demonstrates performance enhancements ranging from 1\% to 7\%. Pipoli et al. \cite{pipoli2022predicting} designed a transformer-based method, Transformer DeepLncLoc, to predict continuous gene expression levels using post-transcriptional information and promoter sequences, addressing the problem as a regression task. The framework of this model uses word2vec embeddings as inputs \cite{zou2019gene2vec}, a positional encoding scheme, and Multi-Headed-Attention layer. In word2vec, the sequences are split into \textit{k}-mer groups of three to form a dictionary of words. The position of the word is tracked by the positional encoding scheme. Their results showed that Transformer DeepLncLoc method produced marginally improved performance compared to existing deep learning technique called Xpresso \cite{agarwal2020predicting} (Mean R\textsuperscript{2} scores: 0.760  versus  0.745). 

Angermueller et al. \cite{angermueller2017deepcpg} developed, DeepCpG, a DL method for the identification of methylation states in single cells. This model contained three modules: DNA module, joint module, and CpG module. In the DNA module, features are extracted from the DNA sequence based on one-hot encoding, which is followed by a CNN model. In the CpG module, a non-linear embedding layer is employed which then becomes an input to a bidirectional gated recurrent unit (BiGRU) network to model correlations between cells. Finally, a joint module combines features obtained from CpG and DNA modules to predict the methylation state at target CpG sites. The DeepCpG method was reported to outperform the existing random forest method (average AUC ROC of 0.89 versus 0.86) to predict methylated versus non-methylated regions. Similarly, Tian et al. \cite{tian2019mrcnn}, designed a DL-based model namely: MRCNN (Methylation Regression Convolutional Neural Networks) for predicting genome-wide DNA methylation status. In this model one-hot-encoding scheme was applied to input data to convert the DNA sequence of length 400 bp into matrices that are fed into the CNN model. The prediction performance of the MRCNN model was evaluated and predicted on two aspects such as binary classification (CpG islands and non-CpG islands), and regression errors (hypomethylation, hypermethylation, and intermediate methylation) performance. The performance of the MRCNN model was reported to be better than that of DeepCpG (AUC ROC of 0.97 versus 0.89). 

Bai et al. \cite{bai2023mlacnn} developed an attention mechanism-based CNN model, MLACNN, to predict genome-wide DNA methylation using WGBS DNA methylation data. The framework of this model consists of a feature encoding scheme and an attention mechanism followed by feature fusion. In the feature encoding scheme three different encoder methods are employed including: nucleotide chemical property coding \cite{chen2017idna4mc}, one-hot encoding \cite{mao2017modeling}, and electron-ion interaction pseudopotentials coding-vector \cite{bai2023mlacnn}. These three feature encodings are fed into three CNN-attention blocks to extract further feature representations. Subsequently, the model applied feature fusion based on the attention mechanism to concatenate the features learned from feature extraction. The model was able to learn features most relevant to the task of methylation prediction. Their results showed that MLACNN model produced better performance compared to existing deep learning techniques called MRCNN and DeepCpG (average AUC ROC of 0.98 versus 0.97 and 0.89). In addition, for a performance comparison of these models see Table \ref{tab2}.

\input{Table2}

\subsection{Prediction of Enhancer-Promoter-interactions}\label{subsec7}
Enhancer–promoter interactions (EPIs) play a central role in the genome by executing transcriptional regulation to control cell differentiation, gene regulation, and disease mechanisms \cite{jing2020prediction, belokopytova2020quantitative,liu2021epihc}. Enhancers regulate the expression patterns of their target genes by interacting directly with their promoter regions \cite{mora2016loop}. The target gene’s expression is controlled by distal regulatory enhancer elements that interact with the proximal promoter regions, and it has been shown that mutations in enhancer regions can change these interactions causing the target gene to be dysregulated \cite{zhang2013chromatin,guo2015crispr}. Diseases such as B-thalassemia and congenital heart disease are caused by mutations in enhancers and promoters, which cause alterations in EPIs \cite{williamson2011enhancers,smemo2012regulatory}. Consequently, there is a significant body of work to develop methods for understanding EP interactions from 1-dimensional genetic and epigenomic marks \cite{whalen2016enhancer,buckle2018polymer}. Broadly these methods consist of two categories of approaches: (i) Physical models that use the knowledge of polymer physics to infer the spatial conformation of regions with EP interactions \cite{buckle2018polymer,chen2016novo,chiariello2016polymer}. (ii) Data-driven and statistical approaches that make use of existing EP-pairs and their interactions to predict if an enhancer and a promoter would interact \cite{whalen2016enhancer,chen2016novo,zeng2018prediction}. The statistical and ML approaches for predicting EPI, unlike the physical model-based methods, have the flexibility of not depending on the choice of the model, and in this paper, we will focus on reviewing ML/DL methods for EPI prediction. 

A seminal work in this regard is the TargetFinder method  \cite{whalen2016enhancer}, which employed boosted trees with functional genomic signals to predict EPIs. Subsequent to this seminal paper, all research groups used the TargetFinder benchmark datasets for training and testing EPI prediction models. 
For instance, Mao et al. \cite{mao2017modeling} designed a predictive method called EPIANN (EPI attention-based neural network) that used sequential features to predict EPIs using DNA sequences. The EPIANN integrates the enhancer and promoter features obtained from convolutional layers with an attention matrix, followed by another set of convolutional layers, concatenation, and a classification layer for predicting of EPIs. By identifying specific regions in promoter and enhancers that drive interactions, the method produces paired attention scores at the sequence level. In terms of AUC ROC, AUC PR, and F1 scores, the EPIANN method was reported to have a slightly better performance than TargetFinder (min AUC ROC of 0.918 versus 0.896 and max AUC ROC of 0.959 versus 0.951) on all six cell lines. Moreover, Singh et al. \cite{singh2019predicting} developed a DL-based architecture called SPEID (Sequence-based Promoter-Enhancer Interaction with Deep learning) and merged the CNN with LSTM to predict EPIs. In this model, first the CNN model was applied to learn hidden informative subsequence-level features in addition to enhancer and promoter sequences, respectively. The next layer constituted a LSTM model, responsible for identifying long-range dependences and for combining the extracted subsequence-level features from the previous layers. The prediction performance of the SPEID model was reported to be better than that of TargetFinder for all six cell lines.

Similarly, Zhuang et al. \cite{zhuang2019simple} developed a simple CNN-based prediction method by simplifying the SPEID method. The key aspect of this model is its simplicity because it uses a single-layer CNN for feature learning. The CNN hyper-parameters had the same default settings as the SPEID method hyper-parameters and produced slightly better or equal prediction performance than SPEID in terms of AUC PR and AUC ROC. The prediction result of the simple CNN model was better as compared to EPIANN on six cell lines (min AUC ROC of 0.941 versus 0.918 and max AUC ROC of 0.962 versus 0.959). Likewise, Hong et al. \cite{hong2020identifying} developed a DL-based method, EPIVAN (EPIs with pre-trained Vector and Attention-based Neural Networks), for the prediction of EPIs using genomic sequences. The EPIVAN model consists of four steps: sequence embedding, feature engineering, attention mechanism, and prediction.  EPIVAN used pre-trained DNA2vec vectors to produce a sequence embedding. Next, it employed a CNN to learn important features from promoters and enhancers datasets, respectively, followed by concatenation which was fed into a bi-directional GRU model (BiGRU). The BiGRU model has two state vectors that read features from both the forward and reverse directions at the same time. Finally, the attention mechanism is added alongside the BiGRU layer to adaptively learn the weights for salient features. The EPIVAN models showed improved prediction results than that of a simple CNN on six cell lines (min AUC ROC of 0.950 versus 0.933 and max AUC ROC of 0.985 versus 0.962).

Roy et al. \cite{roy2015predictive} developed a Regulatory Interaction Prediction for Promoters and Long-range Enhancers (RIPPLE) computational model for understanding the relationship between enhancers and promoters. This method used 3C and 5C chromatin interaction data with minimal regulatory genomic datasets containing 8 histone marks and 15 TF binding sites for five cell lines. RIPPLE showed the potential to produce genome-wide interaction maps and predict interactions in new cell lines. In another study, Jing et al. \cite{jing2020prediction} used CNN and LSTM to extract hidden features and then applied adversarial neural networks with a gradient reversal layer (GRL) to reduce domain-specific features. They reported higher values for AUC ROC, AUC PR, and F1 as compared to the previously published RIPPLE \cite{roy2015predictive} method (min AUC ROC of 0.77 versus 0.61 and max AUC ROC of 0.83 versus 0.68).

Belokopytova et al. \cite{belokopytova2020quantitative} pointed out that the sequences of promoter and enhancer from the same chromosomes have a large level of redundant information and lead to the overestimated prediction performance of the existing EPI models. They randomly selected two chromosomes of the enhancer and promoter pair as a validation dataset and the remaining enhancer and promoter pairs were considered as a training dataset, and showed that the performance got dropped. Liu et al. \cite{liu2021epihc} presented a CNN-based method, EPIHC (EPI based on Hybrid features and Communicative learning), which used hybrid features i.e., genomic features and sequence-derived features, along with a communicative learning module. The communicative learning module retained sequence dependency and promoter-enhancer interaction at the segment level.  The EPIHC method obtained better performance in terms of AUC ROC, AUC PR, and F1 scores than EPIVAN, SPEID, and simple CNN methods for all cell lines. 

In another work, Min et al. \cite{min2021predicting} introduced a DL-based framework, EPI-DLMH (Enhancer Promoter Interactions-Deep Learning Matching Heuristics), that used DNA sequences for the prediction of EPIs. In the EPI-DLMH method, the local features were extracted using a two-layer CNN, while a bidirectional GRU was utilized to capture long-range dependencies among the promoter and enhancer sequences.  In addition, an attention layer was incorporated to calculate the relevance of important features. Then, the learned feature vectors for the enhancer and promoter were appended by using a matched heuristic approach, which employs a set of rules and criteria to find matches in a data structure \cite{boschetti2022matheuristics,min2021predicting}. The prediction outcomes of the EPI-DLMH model were reported to be better than EPIANN on six cell lines (min AUC ROC of 0.948 versus 0.924 and max AUC ROC of 0.977 versus 0.959). Further, Song et al. \cite{song2023deepdualepi} developed a DL-based approach called DeepDualEPI (Deep Dual-channel EPI), for the prediction of EPIs using genomic sequences and genomic signals of four cell lines. The architecture of this approach consists of two modules: the first module uses a two-layer CNN model to extract hidden features from DNA sequences; the second module processes the genomic signals using dilated CNN, BiLSTM, and a Transformer network; the feature maps of both modules are then concatenated to produce hybrid features and output EPI prediction probabilities. The DeepDualEPI models showed improved prediction results than that of Targetfinder on these four cell lines (min AUC ROC of 0.8243 versus 0.7942 and max AUC ROC of 0.9344 versus 0.8671).
Fan et al \cite{fan2022stackepi}, introduced a ML-based model called stackEPI to predict enhancer-promoter interactions from DNA sequences using a stacking ensemble learing techniques. The model merged various encoding methods including PseKNC, Kmer, sequence based information, etc and various ML algorithms including SVM, RF, etc to extract effective information from promoter and enhancer sequences. The prediction outcomes of the StackEPI model were reported to be better than EPIANN on six cell lines (min AUC ROC of 0.937 versus 0.933 and max AUC ROC of 0.990 versus 0.986).
Most recently, Ahmed et al. \cite{ahmed2024epi} developed a transformer-based DL model called EPI-Trans (EPI-Transformer), for the prediction of EPIs using genomic sequences. The architecture of this approach integrates CNN and Transformer to improve the performance of EPI predictions. The CNN module extracts local features from promoter and enhancer sequences; then feature vectors generated by CNN module combined and fed into a transformer module. The transformer module contained positional encoding, Add \& Norm position-wise feedforward network, and multi-head attention layers. The EPI-Tran model obtained better performance in terms of AUC ROC, AUC PR, and F1 scores than simple CNN methods for some cell lines (min AUC ROC of 0.946 versus 0.933 and max AUC ROC of 0.983 versus 0.962).

Overall, while a significant body of work exists for EPI prediction models, one limitation in all the above reviewed papers was highly unbalanced datasets which lead to overfitting and overestimated performance. For further performance comparison of these models see Table \ref{tab3}.  

\input{Table3}

\subsection{Discovery of Chromatin states}\label{subsec8}
Due to DNA packaging and folding of chromosomes, different parts of the genome may interact with each other leading to differential accessibility for transcription factors to bind \cite{rowley2018organizational,bonev2016organization, bickmore2013genome}. As a result, different regions of the genome differ in terms of their potential to get transcribed. At a very broad level, the chromatin can be said to have two states: active (ready for transcription i.e., compartment A), or repressed (compartment B) \cite{rao20143d,lieberman2009comprehensive}. However, there has been work that shows sub-types of these states also referred to as sub-compartments \cite{rao20143d}. In particular, Rao et al. \cite{rao20143d} further refined the A/B compartmental definitions by identifying five Hi-C sub-compartments (A1, A2, B1, B2, B3). Genetic and regulatory information are stored in every human cell chromatin, where DNA is densely packed and wrapped around histone proteins. Gene expression, protein synthesis, biological pathways, and finally complex phenotypes are all affected by chromatin structure. In addition, the methods for accurate detection of chromatin states are also critical to understanding how and when chromatin goes through reorganization and transition from one state to the other. In this section we review deep learning algorithms for finding the chromatin state in genomics sequences on the basis of similarities and differences. 

Zhou and Troyanskaya \cite{zhou2015predicting} developed a method called DeepSEA (deep learning-based sequence analyzer) using CNNs to predict chromatin marks from DNA sequences. This method directly learns a regulatory sequence code from large-scale chromatin profiling data, enabling the prediction of functional elements and variant effects in non-coding regions. Additionally, DeepSEA computes various features for each input variant, such as predicted chromatin effects for histone marks, DNase I hypersensitive sites, TF, and evolutionary conservation scores. The DeepSEA model obtained a median AUC ROC of 0.896, 0.923, and 0.856 on TF binding sites, DNase-I hypersensitive sites, and HM respectively. The prediction outcomes of the DeepSEA model were reported to be better than the previous machine learning gkmSVM model (average AUC ROC of 0.88 versus 0.86). Min et al. \cite{min2017chromatin} introduced a DL model that was a combination of unsupervised representation learning and supervised learning namely, CLSTM (convolutional long short-term memory), for the prediction of chromatin accessible regions from DNA sequences. This model employed a CNN and a bidirectional LSTM with the pre-trained \textit{k}-mer embedding vectors for pattern learning and classification. They employed GloVe (Global Vectors) \cite{pennington2014glove}, an unsupervised learning algorithm, to represent the DNA sequences as word embedding vectors. The results showed that CLSTM model significantly outperformed the existing gkmSVM and DeepSEA models (average AUC ROC of 0.8947 versus 0.866 and 0.887).

In another work, Liu et al. \cite{liu2018chromatin} presented Deopen (Deep openness prediction network), a hybrid computational model based on CNN and \textit{k}-mer features to predict chromatin accessibility.  CNN was used to automatically learn the pattern of DNA sequence, which was then combined with a three-layer feed-forward neural network to learn the high-level representation of \textit{k}-mer spectrum characteristics. The outputs of these two networks were combined, and then fed into a fully connected layer followed by classification. The reported results showed that Deopen model outperformed the existing CLSTM, gkmSVM, and DeepSEA models (average AUC ROC of 0.9086 versus 0.8947, 0.866, and 0.887). The performance of this model is likely attributable to the fact that it uses both conventional features as well as CNN derived features. Hill et al. \cite{hill2023chromdl} developed a deep learning-based architecture, ChromDL, which integrates BiGRU, CNN, and Bidirectional-LSTM units for predicting HM, TF binding sites, and DNase-I hypersensitive sites. ChromDL was shown to be more successful at identifying weak TF binding, which may help define the specificities of TF binding motifs. The reported results showed that ChromDL model was marginally better or in some cases similar in performance to the previous DeepSEA model on TF binding sites (Median AUC ROC of 0.97 versus 0.958) on DNase-I hypersensitive sites (Median AUC ROC of 0.936 versus 0.924), and on HM (Median AUC ROC of 0.864 versus 0.856) using H1-hESC, K562, and HepG2 cells derived from the ENCODE dataset. 

Lanchantin et al. \cite{lanchantin2019graph} developed a graph-based DL method, ChromeGCN, which combined both long-range 3D genome data and the local sequence to predict chromatin state. The CNN model was first employed to find local sequence patterns to discover and learn DNA motifs. It then employed a gated graph convolutional network (GCN) for classification. The performance showed that ChromeGCN model gave slight improvement over the previous CNN model \cite{zhou2018deep} (Mean AUC ROC of 0.909 versus 0.895 and Mean AUC ROC 0.912 versus 0.894). Similarly, Guo et al. \cite{guo2020deepanf} proposed a deep learning method, DeepANF (deep attentive neural framework), to predict chromatin accessibility based on unsupervised Word2Vec embedding representation of DNA sequences. The DeepANF method used CNN and bidirectional GRU (BiGRU) to extract a latent representation of DNA sequences. An attention mechanism was then used to merge the features obtained from CNN and BiGRU to predict chromatin accessibility. The result showed that DeepANF method gave improved performance compared to existing deep learning and machine learning methods (average AUC ROC of 0.919 versus 0.899). 

Farré et al. \cite{farre2018dense} introduced a method based on a dense neural network to predict chromatin state sequence representation of the chromatin structure and chromatin conformation. The sequencing data for a region of a chromosome were used to train the model to predict the appropriate sub-region of the Hi-C contact map (or vice versa). Furthermore, the model was able to solve the inverse problem to produce an optimized 1D sequence annotation of chromatin states that best explain the chromatin conformation. Sensitivity analysis was used to discover the relation between each conformation and a sequence, allowing interpretation of key regulatory features responsible for this relationship, as well as explaining the importance of sequence neighborhood in chromatin structure. Pan et al. \cite{pan2024silencerein} introduced SielenceREIN (Silencers on the Regulatory Element Interaction Network), which utilized the chromatin conformation datasets obtained from ENCODE database for identifying silencers on anchors of chromatin loops. The method utilized a graph-neural network for extracting features based on the GraphSAGE module and subsequently employed CNN to extract feature maps from linear genomic signatures. The feature maps from the CNN and GraphSAGE modules were then concatenated and fed into a MLP classifier to identify silencers.  The results showed that SilenceREIN model outperformed the previous gkmSVM model (AUC ROC of 0.793 versus 0.760).

Ashoor et al. \cite{ashoor2020graph} developed a method namely, SCI (Sub-Compartment Identifier), to predict genomic sub compartments from Hi-C data by applying large-scale information network embedding method \cite{tang2015line} to learn an embedding representation for genomic loci. This was followed by clustering on the learned embeddings.  Finally, a deep neural network was used for classification to predict five sub-compartments (three inactive and two active), each of them having their unique functional and spatial properties. Yang et al. \cite{yang2020cancer} proposed a GAN-based method, ClusterATAC (Cluster Assay for Transposase-Accessible Chromatin), to precisely cluster 401 TCGA tumor samples based on the ATAC-seq data mapped chromatin accessibility profiles. The architecture of ClusterATAC model contained two modules: the Encoder module was based on the GAN framework for model training, and the gaussian mixture model module was used to cluster the results of the encoder module. In the analysis, the 401 TCGA samples were reported to be coming from 22 cancer subtypes. Xiong et al. \cite{xiong2019revealing} used high-coverage Hi-C datasets to introduce a model namely, SNIPER (Subcompartment iNference using Imputed Probabilistic ExpRessions). It divides A/B compartments into A1, A2, B1, B2, and B3 subgroups, which demonstrate association with both genomic and epigenomic features. Two distinct neural network frameworks were used in this computational method: a MLP classifier that classifies the regions into one of five main subcompartment classes, and a denoising autoencoder that extracts features while reducing the dimensionality of the input data.  For a more detailed performance comparison of these models see Table \ref{tab4}.

\input{Table4}

\subsection{Representation learning for epigenetic problems}\label{subsec9}
One of the key factors behind the success of modern AI and deep learning models is their capacity to learn useful and efficient representations \cite{bengio2013representation,bottou2007large,zhong2016overview}.  When a model is trained for a particular task, the step of feature extraction may not be explicit, yet, what is fed into the last classification layer of a neural network can often be viewed as a feature representation \cite{zhong2016overview,yi2019acp}.  At a higher level, there are two widely used paradigms for representation learning. The first among them is the unsupervised paradigm which is based on training an autoencoder on a large dataset of unlabeled examples \cite{lipton2015critical}. The output of the encoder (the latent space) can then be used for downstream supervised tasks. Glimpses of this approach could also be seen in the previous section on detecting chromatin states. The second approach, based on supervised learning, relies on reusing the knowledge of a pre-trained model by training it for tasks which it was not originally trained for. This method is also referred to as transfer learning \cite{weiss2016survey,cook2013transfer,feuz2015transfer}. More concretely, when a classifier is trained for a specific task, the output of the second to last layer can be considered as a feature representation, and the neural network up until that point can be used as a feature extractor and may be fine-tuned for a different task \cite{cook2013transfer,bunrit2020improving}. In this section, we review representation learning methods for epigenetic problems. 

Zhou et al.\cite{zhou2020imputing} developed a method called TDimpute that used DNA methylation data and employed transfer learning with DNN to impute the missing gene expression value. Initially, a model was trained on a pan-cancer dataset. Later, transfer learning was used to adapt it to target cancer types \cite{troyanskaya2001missing}. Schwessinger et al. \cite{schwessinger2020deepc} designed a DL-based architecture called DeepC to predict genome folding from DNA sequence. The DeepC model employed transfer learning for feature extraction and deep neural network for classification. DeepC was trained in two phases. Firstly, the model was trained to predict epigenetic features, using convolution layers to extract hidden features and capture patterns in sequences related to histone modifications and transcription factors. Only the learned feature vectors obtained in the first phase are further transferred to the second phase of the convolution layers i.e., feature extraction modules where they were refined. A stack of dilated CNNs was employed after the feature extraction module to predict the chromatin interaction between 5 kb genomic bins in 1 Mb areas.  

Levy et al. \cite{levy2020methylnet} introduced a DL-based method namely MethylNet for pan-cancer prediction and classification using transfer learning. MethylNet was developed for automatically creating embedding, producing new data, making predictions, and identifying previously unrecognised disease heterogeneity. In this framework, first, the deep learning model was pre-trained with VAE to extract hidden features for unsupervised clustering and dimensionality reduction of the methylation data. Then, the framework incorporated prediction layers to further optimize the encoder for regression, classification, and multi-output regression tasks. Finally, they employed a hyper-parameter scanning method for the prediction layers and feature extraction network to optimize the model parameters. Two methods were used to interpret predictions from MethylNet: (i) SHAP (SHapley Additive ExPlanation) method \cite{lyu2019advances} to predict key methylation states in different cancer subtypes and cell types, and (ii) Comparison of the learned clusters of methylation samples embedding for biological validity. Based on the results, the MethylNet method was reported to have outperformed other machine learning methods in the accuracy of pan-cancer prediction and classification in methylation data (accuracy 0.97 versus 0.84). 

Li et al. \cite{lai2021predicting} presented a deep transfer learning-based method called MetaChrom for predicting the impacts of DNA variations in various cellular contexts, including neurodevelopment and genome-wide epigenomic profiles. This model contained two modules: first, a sequence model based on the ResNet architecture, namely a sequence encoder that is designed to extract cell-type-specific features directly from the DNA sequence. After being pre-trained on large public datasets, the second module contains a CNN architecture, namely a meta-feature extractor to extract hidden features from DNA sequences. Subsequently, the feature maps from both modules were integrated to predict epigenetic profiles. The objective of this model is to better understand how genetic variation may influence epigenetic regulation and gene expression during important stages of brain formation. The reported results showed that MetaChrom model gave improved performance to the previous DanQ \cite{quang2016danq} model (Average AUC ROC of 0.89 versus 0.86).
Li et al. \cite{li2023epiteamdna} developed a predictive model named EpiTEAmDNA that utilised transfer learning and ensemble learning techniques to improve the representation of sequence features to predict different DNA epigenetic modifications across 15 species. In this study, 14 various feature extraction techniques, namely, k-mer, nucleotide chemical properties, and so on, were used to extract hidden features from DNA sequences, and then eight different ML methods, namely random forest, adaboost, etc., were applied to these feature extraction methods. Similarly, it employed a CNN to learn important features from DNA sequences, followed by concatenation, which integrates the feature vector obtained from the ML and DL baseline models and then fed into a logistic regression for classification. The EpiTEAmDNA models showed improved prediction results on 27 datasets (min ACC of 0.7592, max ACC of 0.9906, and avg ACC of 0.8810). 

Wang et al.\cite{wang2024bert} introduced a DL-based method called BERT-TFBS (Bidirectional Encoder Representations from Transformers-Transcription Factor Binding Sites) for predicting TF binding sites from DNA sequences. The BERT-TFBS model integrates a pre-trained BERT model namely DNABERT-2, with a CNN and a convolutional block attention module (CBAM), and an output module. The model used transfer learning by employing the pre-trained DNABERT-2 model to capture intricate long-term dependencies in DNA sequences. After that, high-order local features were extracted by the CNN and CBAM modules. The output module employed a MLP together with the learnt sequence features to predict TFBSs in the DNA sequences. A fully connected layer with dropout and a fully connected layer with the SoftMax function made up the two layers of the MLP. The result showed that the BERT-TFBS method gave improved performance compared to existing deep learning methods (AUC ROC of 0.919 versus 0.887).
Salvatore et al. \cite{salvatore2023transfer} introduced a DL-based transfer learning technique called ChromTransfer that fine-tunes models for predicting cell-type-specific chromatin accessibility. The model applied a pre-trained, cell-type-agnostic model of open chromatin regions to enhance the prediction performance of six cell lines. Insights into the regulatory code are obtained by using this method to identify sequence features that match binding site sequences of important TFs for prediction. The prediction results of the ChromTransfer model obtained an AUC ROC between 0.79 and 0.89 for all six cell lines.

Wang et al. \cite{wang2022imputing} proposed a transfer learning-based neural network method, TDimpute-DNAmeth, to impute the missing values of DNA methylation data. In this method, first, the original benchmark dataset was split into two parts i.e., target dataset and pan-cancer dataset. Using the pan-cancer dataset, a generic imputation model was first built for all the cancer types. Then, transfer learning was used to fine-tune the model for the target cancer type. Based on the results, the TDimpute-DNAmeth method outperformed other methods on independent datasets. Chen et al. \cite{chen2022exploiting} proposed a deep transfer learning method, TLVar (Transfer Learning Variants), to predict functional non-coding variants (NCVs) using flanking genomic sequences. The CNN used in the deep transfer learning model includes two convolutional and dense layers. In this framework, the CNN used large-scale generic functional NCVs such as ORegAnno \cite{lesurf2016oreganno}, ClinVar \cite{landrum2016clinvar}, and HGMD \cite{stenson2020human} to pre-train a base network. Then, the target network was fine-tuned by retraining only the dense layers using context-specific functional NCVs while the convolutional layers were transferred without re-training. Finally, they produced binary values to predict functional or non-functional NCVs. Their TLVar model produced better results than other models (AUC ROC is 0.634 versus 0.612 and 0.695 versus 0.685). For a performance comparison of these models see Table \ref{tab5}.

\input{Table5}

\section{Challenges and Recommendations}\label{sec5}
In this section, we will present several challenges that we have identified after reviewing a diverse body of work related to AI methods available in the literature for solving problems pertaining to epigenetic data. We will also provide recommendations for addressing these challenges.

One common problem with the data used in the reviewed studies is that most of the datasets happen to be considerably imbalanced with respect to the variable that the models are supposed to predict. For instance, a frequent scenario could be that in a dataset collected to study gene expression, the number of examples in which a gene was expressed could be outnumbered significantly by those in which the gene was repressed \cite{kundaje2015integrative}. Although this imbalance can be a consequence of the inherent biology, for AI models, different distributions of the ground-truth variable can pose serious difficulties. While every effort should be made to address this issue at the stage of data collection, more often than not the nature of the data remains inherently imbalanced. Moreover, AI researchers usually focus on algorithm development relying on publicly available datasets which have already been collected by other research groups. Consequently, given an imbalanced dataset, as an AI researcher, a number of steps should be considered at every level ranging from data-preparation, data augmentation, to the selection of loss-function and learning paradigm as well as training parameters. 

In this regard, firstly, data augmentation techniques need to be thoroughly revisited. While there are well-known standard techniques for data augmentation as applied to image data (e.g., geometric, scale, and intensity transforms), such intuitive methods do not scale up for augmenting genomic data. As a result, generative models such as GANs, and diffusion-based transformer networks should be considered and further developed for genomic sequence augmentation \cite{kircher2022augmentation}. From the perspective of learning paradigms, contrastive techniques such as supervised contrastive learning \cite{khosla2020supervised} can be very useful for building predictive models using imbalanced datasets. Contrastive methods can learn representations that allow maximizing distances between examples from different classes while also minimizing distances between examples from the same class. This is often done by forming triplets of examples which enables having multiple triplets for each of the limited number of data points for the minority class. Techniques such as few shot, one shot, and zero shot learning \cite{chen2023zero,kadam2020review,rahman2018unified} also need to be explored and enhanced for genomic contexts. Further, during the actual training process, it is important to ensure that the training batches are also sampled in a balanced way, so that while optimization, gradients are computed based on examples from all the classes.

In representation learning for downstream tasks, achieving optimal validation performance is often challenging due to insufficient attention given to critical elements of the validation process such as data splitting ratio, feature selection, early stopping, and hyperparameter tuning. To enhance performance on unseen data, it is essential to meticulously consider and precisely describe the validation procedure, encompassing data stratified splitting, well-defined feature selection criteria, effective hyperparameter tuning, and early stopping to prevent overfitting. Additionally, transfer learning, ensembling, model interpretability using techniques such as attention mechanisms \cite{ashurov2023improved} and regularization techniques \cite{barone2017regularization} should be incorporated.

Harmonizing models trained on multiple datasets and repositories without retraining poses a significant challenge due to differences in data distributions, feature representations, and model architectures. These disparities can lead to inconsistencies and suboptimal performance when combining their predictions directly. To address this, transfer learning techniques can be employed by fine-tuning the models on a common task or dataset, allowing them to share and adapt their learned representations. Model ensembling, through techniques like weighted averaging or stacking, can also be used to combine outputs and capture diverse viewpoints, resulting in more robust predictions that leverage the strengths of each model. More importantly, methods for domain transfer and dataset bias unlearning \cite{tommasi2017deeper,ashraf2018learning} should be employed to improve cross-dataset generalization. 

Lastly, the actual deployability of computational models, particularly in the context of deep learning applied to epigenetics, will continue to remain a challenge unless more structured validation techniques are introduced. In this regard, it is paramount to develop wet experimental validation protocols for testing the prediction of AI models on novel data and then verifying the prediction by means of assessing concordance with the outcome of wet experiments. The experimental outcome can then be used to adapt the AI models such that the models can be improved by following principles of continuous-learning-AI \cite{elemento2021artificial,hassabis2017neuroscience}.

\section{Conclusion}
With the advent of high-throughput sequencing, the field of epigenetic sequence analysis stands at the interface of computational biology and machine learning, offering the promise of furthering our understanding of gene regulation. The audience of this review is both AI researchers and epigeneticists. We have provided a taxonomy of epigenetic sequence analysis problems that are approachable through AI-based methodology to help the AI researchers to find new and challenging problems which are good candidates to be solved through AI. We then map the above problems to the published research that has employed AI models to approach them. In so doing, we have reviewed and described a spectrum of deep learning architectures employed in analyzing epigenomic data, highlighting their strengths, limitations, and potential applications. As we navigated through the nuanced challenges of epigenetic sequence analysis, it became evident that a comprehensive approach demands an understanding of both the biological mechanisms and AI computational algorithms. The integration of deep learning architectures has paved the way for significant advancements in predicting functional elements, deciphering regulatory mechanisms, and enhancing our grasp of gene expression patterns. However, this expedition is not without its obstacles; these hurdles encompass diverse aspects, including addressing imbalanced datasets to ensure learning of useful representations, embracing a wide array of performance metrics, enhancing model interpretability, improving data harmonization strategies, and refining validation protocols for assessing the predictions of AI models through outcomes of wet experiments. To overcome these challenges and to tap into the full potential of AI for epigenetic sequence analysis, collaborative efforts across biology, data science, and machine learning are essential. A concerted approach that combines domain expertise with innovative algorithmic solutions will catalyze breakthroughs in our understanding of epigenetic regulation. In the last section of this review, we have described and identified the above challenges and have provided several recommendations and ideas on how to address these issues.

\section*{Declaration of competing interest}
No competing interest is declared.

\section*{CRediT authorship contribution statement}
\textbf{Muhammad Tahir:} conceived the idea, analyzed the studies and wrote this manuscript.
\textbf{Mahboobeh Norouzi:} analyzed the studies and wrote this manuscript.
\textbf{Shehroz S. Khan:} conceived the idea, analyzed the studies and wrote this manuscript.
\textbf{James R. Davie:} conceived the idea, analyzed the studies and wrote this manuscript.
\textbf{Soichiro Yamanaka:} conceived the idea, analyzed the studies and wrote this manuscript.
\textbf{Ahmed Ashraf:} conceived the idea, analyzed the studies and wrote this manuscript. All authors read and approved the final manuscript.

\section*{Acknowledgments}
Financial support from the following funding agencies is acknowledged: 
• Canadian Institutes of Health Research (CIHR) 
• Japan Agency for Medical Research and Development (AMED)
\bibliographystyle{unsrt}

\bibliography{cas-refs}


\end{document}

%% file: Table1.tex
\begin{table*}[t]

\caption{Summary of the literature based on DL approaches used for Predicting/Detecting Disease Markers.\label{tab1}}
\tabcolsep=0pt
\begin{tabular*}{\textwidth}{@{\extracolsep{\fill}}>{\raggedright\arraybackslash}p{2.5cm}>{\raggedright\arraybackslash}p{2.5cm}>{\raggedright\arraybackslash}p{2.5cm}>{\raggedright\arraybackslash}p{2.5cm}>{\raggedright\arraybackslash}p{2.5cm}>{\raggedright\arraybackslash}p{2.5cm}@{\extracolsep{\fill}}}
\toprule%
Authors/Refs.   &   Model’s Name   &   DL approaches   &   Dataset(s) (Input)   &   Results (Output)   &   Performance   \\

\midrule

Li et al.\cite{li2021dismir} &   DISMIR   &   -CNN\newline-LSTM   &   DNA sequences with methylation state   &   Early detection    &   -AUC ROC = 0.9969 \newline-AUC ROC = 0.9112   \\

Liu et al.\cite{liu2019dna} &   DNA methylation markers via DL   &   Fully Connected Deep Networks   &   Promoter markers \&  CpG markers   &   Pan-cancer   &   AUC ROC = 0.995 \newline-AUC ROC = 0.993   \\

Albaradei et al.\cite{albaradei2021metacancer} &   MetaCancer   &   DNN   &   -DNA methylation and microRNA sequencing \newline -RNA-Seq   &   Pan-cancer metastasis   &   ACC = 88.85\%   \\

Xiao et al.\cite{xiao2021cancer} &   WGAN Model   &   -DNN \newline-GAN   &   \text TCGA-Seq &   Cancer Diagnosis    &   ACC=98.33\%, 96.67\%, \& 96.67\%   \\

Rajpal et at. \cite{rajpal2023xai} &   XAI-MethylMarker   &  Feed-forward neural network   &   DNA methylation   &   Biomarker for breast cancer classification   &   ACC = 0.8145   \\

Jiang et al.\cite{jiang2020generative} &   GAN-DAEMLP   &   GAN   &   RNA-seq   &   Gene disease    &   AUC ROC =0.6700   \\

Zhang et al. \cite{zhang2021omiembed} &   OmiEmbed   &   CNN   &   -DNA methylation \newline-miRNA expression \newline-Gene expression    &   Cancer survival predication \newline-tumor classification etc   &  AUC ROC = 0.9943   \\

Zhang et al.\cite{zhang2022transformer} &   T-GEM   &   Transformer   &   Gene expression and RNA-Seq   &   -Cancer type prediction \newline-Immune cell type classification   &   ACC = 90.73\%   \\

Yin el al. \cite{yin2019deephistone} &   DeepHistone   &   CNN    &   -DNA sequences \newline-DNase-Seq    &   Histone markers   &   AUC ROC = 0.9065   \\

Baisya et al.\cite{baisya2020prediction} &   DeepPTM   &   Neural Networks   &   DNA sequences and TF-binding data   &   Histone markers   &   AUC ROC =0.9543   \\

Li et at.\cite{li2022identifying} &   iHMnBS   &   -CNN \newline -GRU   &   -DNA sequences \newline-DNase-seq   &   -HMs markers \newline-Binding sites   &   AUC ROC = 0.9411   \\

\bottomrule
\end{tabular*}

\end{table*}

%% file: Table2.tex
\begin{table*}[t]

\caption{Summary of the literature based on DL approaches used for Gene Expression Prediction  and DNA methylation.\label{tab2}}
\tabcolsep=0pt
\begin{tabular*}{\textwidth}{@{\extracolsep{\fill}}>{\raggedright\arraybackslash}p{2.5cm}>{\raggedright\arraybackslash}p{2.5cm}>{\raggedright\arraybackslash}p{2.5cm}>{\raggedright\arraybackslash}p{2.5cm}>{\raggedright\arraybackslash}p{2.5cm}>{\raggedright\arraybackslash}p{2.5cm}@{\extracolsep{\fill}}}
\toprule%
Authors/Refs.	&	Model’s Name	&		DL approaches	&	Dataset(s) (Input)	&	Results (Output)	&	Performance	\\

\midrule
Singh et al. \cite{singh2016deepchrome} &	DeepChrome	&		CNN	&	HMs (ChIP-seq) 	&	Gene expression	&	AUC ROC = 0.8008	\\

Singh et al. \cite{singh2017attend} &	AttentiveChrome	&		LSTM	&	HMs (ChIP-seq) 	&	Gene expression	&	AUC ROC = 0.8133	\\

Sekhon et al. \cite{sekhon2018deepdiff} &	DeepDiff	&		LSTM	&	HMs (ChIP-seq) 	&	Gene expression	&	- - -	\\

Cheng et al.\cite{cheng2020simplechrome} &	SimpleChrome	&		MLP	&	HMs (ChIP-seq) 	&	Gene expression	&	AUC ROC = 0.809	\\

Frasca et al.\cite{frasca2022accurate} &	ShallowChrome	&		Logistic Regression	&	HMs (ChIP-seq) 	&	Gene expression	&	AUC ROC = 0.8737	\\

Hamdy et al.\cite{hamdy2022convchrome} &	ConvChrome	&		-CNN \newline-Self-attention mechanism 	&	HMs (ChIP-seq)	&	Gene expression	&	AUC ROC = 0.8399 	\\

Chen et al.\cite{chen2022predicting} &	TransferChrome	&		-CNN \newline-Self-attention mechanism 	&	HMs (ChIP-seq)	&	Gene expression	&	AUC ROC = 0.8479 	\\

Hamdy et al.\cite{hamdy2023deepepi} &	DeepEpi		&		-CNN \newline-LSTM \newline-Self-attention mechanism 	&	HMs (ChIP-seq)	&	Gene expression	&	AUC ROC = 0.8887 	\\

Tahir et al.\cite{Tahir2024TransformerChrome} &	TransformerChrome		&		Transformer 	&	HMs (ChIP-seq)	&	Gene expression	&	AUC ROC = 0.8152 	 \\

Pipoli et at.\cite{pipoli2022predicting} &	Transformer DeepLncLoc	&		Transformer 	&	Continuous gene expression levels	&	Post-transcriptional information and DNA sequences	& Mean $R^2$: 0.760	\\

Angermueller et al.\cite{angermueller2017deepcpg} &	DeepCpG	&		-RNN\newline-CNN	&	Methylation (DNA sequence and features)	&	DNA methylation	&	AUC ROC = 0.89	\\

Tian et al.\cite{tian2019mrcnn} &	MRCNN	&		CNN	&	DNA sequences	&	DNA methylation	&	AUC ROC = 0.97	\\

Bai et al.\cite{bai2023mlacnn} &	MLACNN	&		-CNN \newline-Attention mechanism 	&	DNA sequences	&	DNA sequences	&	AUC ROC = 0.98\\

\bottomrule
\end{tabular*}
\end{table*}

%% file: Table3.tex
\begin{table*}[b]

\caption{Summary of the literature based on DL approaches used for Enhancer-Promotor-interaction prediction.\label{tab3}}
\tabcolsep=0pt
\begin{tabular*}{\textwidth}{@{\extracolsep{\fill}}>{\raggedright\arraybackslash}p{2.5cm}>{\raggedright\arraybackslash}p{2.5cm}>{\raggedright\arraybackslash}p{2.5cm}>{\raggedright\arraybackslash}p{2.5cm}>{\raggedright\arraybackslash}p{1.0cm}>{\raggedright\arraybackslash}p{3.5cm}@{\extracolsep{\fill}}}
\toprule%

Authors/Refs.   &   Model’s Name   & DL approaches   & Dataset(s) \newline(Input)   & Results\newline (Output)   & Performance   \\
\midrule

Whalen et al. \cite{whalen2016enhancer} &   TargetFinder   & Boosted Trees   & DNA Sequences   & EPIs   & -Min AUC ROC = 0.903 \newline-Max AUC ROC = 0.951   \\

Mao et al. \cite{mao2017modeling} &   EPIANN   & CNN   & DNA Sequences   & EPIs   & -Min AUC ROC = 0.918 \newline-Max AUC ROC = 0.959   \\

Singh et al.\cite{singh2019predicting} &   SPEID   & -CNN \newline-LSTM   & DNA Sequences   & EPIs   & -Min AUC ROC = 0.904 \newline-Max AUC ROC = 0.950   \\

Zhuang et al.\cite{zhuang2019simple} &   SIMCNN   & -CNN \newline-Transfer learning   & DNA Sequences   & EPIs   & -Min AUC ROC = 0.933 \newline-Max AUC ROC = 0.962   \\

Hong et al.\cite{hong2020identifying} &   EPIVAN   & -CNN \newline-BiGRU   & DNA Sequences   & EPIs   & -Min AUC ROC = 0.950 \newline-Max AUC ROC = 0.985   \\

Jing et al.\cite{jing2020prediction} &   SEPT   & -CNN \newline-LSTM   & DNA Sequences   & EPIs   & - - -   \\

Liu et al.\cite{liu2021epihc} &   EPIHC   & CNN   & DNA Sequences   & EPIs   & -Min AUC ROC = 0.910 \newline-Max AUC ROC = 0.955   \\

Min et al. \cite{min2021predicting} &   EPI-DLMH   & -CNN \newline -BiGRU   & DNA Sequences   & EPIs   & -Min AUC ROC = 0.948 \newline-Max AUC ROC = 0.977  \\

Fan et al. \cite{fan2022stackepi} &   StackEPI  & MLP   & DNA Sequences   & EPIs   & -Min AUC ROC = 0.937 \newline-Max AUC ROC = 0.990  \\

Song et al. \cite{song2023deepdualepi} &   DeepDualEPI   & -CNN \newline -BiLSTM \newline-Transformer   & -DNA Sequences \newline-Genomic signals    & EPIs   & -Min AUC ROC = 0.824 \newline-Max AUC ROC = 0.934   \\

Ahmed et al. \cite{ahmed2024epi} &   EPI-Tran   & -CNN \newline-Transformer   & DNA Sequences    & EPIs   & -Min AUC ROC = 0.946 \newline-Max AUC ROC = 0.983   \\

\bottomrule
\end{tabular*}
\end{table*}

%% file: Table4.tex
\begin{table*}[t]

\caption{Summary of the literature based on DL approaches used for prediction of Chromatin states and subtype discovery.\label{tab4}}
\tabcolsep=0pt
\begin{tabular*}{\textwidth}{@{\extracolsep{\fill}}>{\raggedright\arraybackslash}p{2.5cm}>{\raggedright\arraybackslash}p{2.5cm}>{\raggedright\arraybackslash}p{2.5cm}>{\raggedright\arraybackslash}p{2.5cm}>{\raggedright\arraybackslash}p{2.5cm}>{\raggedright\arraybackslash}p{4.0cm}@{\extracolsep{\fill}}}
\toprule%

Authors/Refs. & Model’s Name & DL approaches & Dataset(s) (Input) & Results (Output) & Performance \\

\midrule
Zhou and Troyanskaya\cite{zhou2015predicting} & DeepSEA & -CNN & DNA sequences  & -Chromatin accessible region & TF binding sites (median AUC ROC = 0.896) \newline -DNase-I hypertensive sites (median AUC ROC = 0.923) \newline -HM (median AUC ROC = 0.856)\\

Min et al.\cite{min2017chromatin} & CLSTM & -CNN \newline -LSTM & DNA sequences  & Chromatin accessible region & Avg: AUC ROC = 0.8947 \\

Liu et al.\cite{liu2018chromatin} & Deopen & -CNN & DNA sequences  & Chromatin accessible region & AUC ROC = 0.9086 \\

Lanchantin et al.\cite{lanchantin2019graph} & ChromGCN & -CNN \newline-GCN & DNA sequences and 3D genome data & Chromatin accessible region & -AUC ROC=0.909 \newline-AUC ROC=0.912 \\

Guo et al.\cite{guo2020deepanf} & DeepANF & -CNN \newline-BiGRU & DNA sequences  & Chromatin accessible region & Avg: AUC ROC = 0.919 \\

Hill et al.\cite{hill2023chromdl} & ChromDL & -CNN \newline-BiGRU \newline-BiLSTM & DNA sequences  & Chromatin accessible region & -TF binding sites \newline(AUC ROC = 0.97) \newline-DNase-I hypertensive sites (AUC ROC = 0.936) \newline-HM (AUC ROC = 0.864) \\

Pan et al.\cite{pan2024silencerein} & SilenceREIN & -CNN \newline-Graph Neural Network & DNA sequences  & Silencers on anchors of chromatin loops & AUC ROC = 0.793 \\

Farré et al.\cite{farre2018dense} & DL-based Model & Dense Neural network & -DNA Sequence \newline-DNase-I hypersensitive signals (DHSs) & Contact map  & - - - \\

Ashoor et al.\cite{ashoor2020graph} & SCI & DNN & -DNA sequence \newline-Genomic Bins & Sub-compartments & - - - \\

Yang et al.\cite{yang2020cancer} & ClusterATAC & GAN & ATAC-seq & Pan-cancer & - - - \\

\bottomrule
\end{tabular*}
\end{table*}

%% file: Table5.tex
\begin{table*}[t]

\caption{Summary of the literature based on DL approaches used for Representation learning for epigenetic problems.\label{tab5}}
\tabcolsep=0pt
\begin{tabular*}{\textwidth}{@{\extracolsep{\fill}}>{\raggedright\arraybackslash}p{2.5cm}>{\raggedright\arraybackslash}p{2.0cm}>{\raggedright\arraybackslash}p{2.5cm}>{\raggedright\arraybackslash}p{2.5cm}>{\raggedright\arraybackslash}p{3.5cm}>{\raggedright\arraybackslash}p{2.5cm}@{\extracolsep{\fill}}}
\toprule%

Authors/Refs. & Model’s Name & DL approaches & Dataset(s) (Input) & Results (Output) & Performance \\
\midrule
Zhou et al.\cite{zhou2020imputing} & TDimpute & Transfer learning & DNA sequences  & Pan-cancer & PR-AUC from 0.601 to 0.983 \\

Schwessinger et al.\cite{schwessinger2020deepc} & DeepC & -Transfer learning \newline-CNN & DNA sequences  & Pan-cancer & - - - \\

Wang et al.\cite{wang2022imputing} & TDimpute-DNAmeth & Transfer learning  & DNA methylation & Unknown-cancer type & - - - \\

Levy et al.\cite{levy2020methylnet} & MethylNet & -Transfer learning \newline-deep learning (NN) & DNA methylation & Pan-cancer & ACC=0.97 \\

Li et al.\cite{li2023epiteamdna} & EpiTEAmDNA & -Transfer learning \newline-ML(RF, AB, etc) \newline-CNN & DNA Sequences & DNA methylation & -Min ACC=0.7592 \newline-Max ACC=0.9906 \newline-Avg: ACC=0.8810 \\

Li et al.\cite{lai2021predicting} & MetaChrom  & -Transfer learning \newline-CNN \newline-ResNet & DNA sequences &  Epigenomic Profile & Avg: AUC ROC =0.89 \\

Salvatore et al. \cite{salvatore2023transfer} &   ChromTransfer   & -Transfer learning \newline-CNN   & DNA Sequences   & chromatin accessibility   & -Min AUC ROC = 0.79 \newline-Max AUC ROC = 0.89   \\

Wang et al.\cite{wang2024bert} & BERT-TFBS & -Transfer learning \newline-CNN \newline-MLP & DNA sequences &  TFBSs & AUC ROC =0.919 \\

Chen et al.\cite{chen2022exploiting} & TLVar & -Transfer learning \newline-CNN & DNA sequences &  NCVs & -AUC ROC =0.634 \newline-AUC ROC =0.685 \\

\bottomrule
\end{tabular*}
\end{table*}